\documentclass[preprint]{aastex}
\usepackage{natbib}
\tighten

\shortauthors{BAUER, CONDON, THUAN \& BRODERICK}
\shorttitle{X-RAY IDS}

\begin{document}

\title{\bf RBSC-NVSS SAMPLE. I. RADIO AND OPTICAL IDENTIFICATIONS OF A
COMPLETE SAMPLE OF 1500 BRIGHT X-RAY SOURCES} 

\author{Franz E.~Bauer\altaffilmark{1}}

\affil{National Radio Astronomy Observatory\altaffilmark{2} \\
Department of Astronomy, University of Virginia, \\
520 Edgemont Road, Charlottesville, VA 22903 \\
E-mail: fbauer@nrao.edu}

\author{J.~J.~Condon}

\affil{National Radio Astronomy Observatory\altaffilmark{2} \\
520 Edgemont Road, Charlottesville, VA 22903 \\
E-mail: jcondon@nrao.edu}

\author{Trinh X. Thuan}

\affil{Department of Astronomy, University of Virginia, \\
P.O. Box 3818, Charlottesville, VA 22903-3818 \\
E-mail: txt@virginia.edu}

\and

\author{J. J. Broderick}

\affil{Department of Physics, Virginia Polytechnic Institute and State
University \\
Blacksburg, Virginia 24061 \\
E-mail: jjb@vt.edu}

\altaffiltext{1}{Visiting Astronomer, Kitt Peak National Optical Observatories, operated
by the Association of Universities for Research in Astronomy, Inc., under 
contract with the National Science Foundation}
\altaffiltext{2}{The National Radio Astronomy Observatory is a
facility of the National Science Foundation operated under cooperative
agreement by Associated Universities, Inc.}

\begin{abstract}
We cross-identified the {\it ROSAT} Bright Source Catalog (RBSC) and
the NRAO VLA Sky Survey (NVSS) to construct the RBSC-NVSS sample of
the brightest X-ray sources ($\geq 0.1$ counts s$^{-1} \sim 10^{-12}$
ergs cm$^{-2}$ s$^{-1}$ in the 0.1--2.4 keV band) that are also radio
sources ($S \geq 2.5$ mJy at 1.4 GHz) in the 7.8 sr of extragalactic
sky with $|b| > 15^\circ$ and $\delta > -40^\circ$. The sky density of
NVSS sources is low enough that they can be reliably identified with
RBSC sources having rms positional uncertainties $\geq 10''$. We used
the more accurate radio positions to make reliable 
X-ray/radio/optical identifications down to the POSS plate limits. We
obtained optical spectra for many of the bright identifications lacking
published redshifts. The resulting X-ray/radio sample is unique in its
size ($N \sim 1500$ objects), composition (a mixture of nearly normal
galaxies, Seyfert galaxies, quasars, and clusters), and low average
redshift ($\langle z\rangle \sim 0.1$).
\end{abstract}


\vfill\eject

\section{INTRODUCTION}

The {\it ROSAT} All-Sky Survey Bright Source Catalogue
\citep[][RBSC revision 1RXS]{voges99} contains the first large all-sky
sample of the brightest X-ray sources, analogous in many respects to
the optical NGC catalog. It was derived from the soft (0.1--2.4 keV)
X-ray survey performed during the first half year of the {\it ROSAT}
mission in 1990/91. The catalog sky coverage is 92\%, and there are
8,547 sources above its 0.1 counts s$^{-1}$ ($\sim 10^{-12}$ ergs
s$^{-1}$ cm$^{-2}$) completeness limit. \citet{bade98} found that
about one third of the RBSC sources can be reliably identified with
galactic stars, while most of the rest are extragalactic. The extragalactic
content of the RBSC comprises a diverse blend of (1) normal spiral
galaxies whose X-ray emission is powered by stars and stellar
remnants, (2) elliptical galaxies with hot gaseous halos, (3) AGN in
Seyfert galaxies, elliptical galaxies, quasars, and BL Lac objects,
and (4) clusters of galaxies. The large number of sources in this
catalog easily permits statistical analyses of {\it each} type of
X-ray object. However, the essential properties of these X-ray sources
cannot be determined from the X-ray data alone---we need observations
in optical and other wavebands to measure their distances, identify
their energy sources, etc. Such observations are possible only for
those RBSC sources whose optical counterparts have been identified. In
this paper we present reliable radio and optical identifications for
sources in the RBSC complete sample.

Most RBSC sources have rms positional uncertainties $\geq 10''$ and
the sky density of faint optical objects is high, so only the nearest
extragalactic X-ray sources can be optically identified by position
coincidence alone \citep[cf.][]{bade98}. Fortunately, most
extragalactic RBSC sources are also radio sources in the 1.4 GHz
NRAO/VLA Sky Survey \citep[][NVSS]{condon98a}, whose sky density is
low enough for identification with RBSC sources. Since the radio
positions are significantly more accurate, the radio sources may be
optically identified, yielding optical identifications for the
corresponding X-ray sources as well. The NVSS covers the 10.3 sr of
sky north of $\delta = -40^\circ$ and contains over $1.8
\times 10^6$ sources stronger than its 2.5 mJy beam$^{-1}$
completeness limit. Since the NVSS was made with relatively low
resolution ($45\arcsec$ FWHM), it does not discriminate against
moderately extended radio sources in nearby galaxies and clusters. Its
rms positional uncertainties range from $< 1\arcsec$ for the $N
\approx 4 \times 10^5$ sources stronger than 15 mJy to 7$\arcsec$ for
the faintest (S = 2.3 mJy) detectable sources, allowing us to make
optical identifications with objects as faint as R $\approx
21$.

We optically identified the RBSC-NVSS sources with objects in the
United States Naval Observatory catalog A2.0 \citep[][USNO]{monet98}.
The USNO catalog contains 526,280,881 objects detected by the
Precision Measuring Machine on the Palomar Optical Sky Survey I
(POSS-I) blue O and red E plates, the UK Science Research Council
SRC-J survey plates, and the European Southern Observatory ESO-R
survey plates. The catalog was compiled from the blue/red overlaps
(within 2$\arcsec$) of the detection lists generated from scans of
POSS-I O and E plates centered on $\delta > -18^\circ$ and SRC-J and
ESO-R plates centered on $\delta < -20^\circ$. The stated astrometric
and photometric errors are about $0\farcs25$ and 0.5 mag rms,
respectively. The USNO catalog covers the entire sky and probes as deep as
B=21 (O plates), R=20 (E plates), J=22, and F=21 for objects with
appropriate colors. 

Section~\ref{cid} explains our method for making the identifications
and assessing their reliabilities.  The results are presented in
Section~\ref{results}.

\section{Cross-identifications}\label{cid}

In this section we present a sequence of increasingly powerful methods
for making cross-identifications and directly evaluating the their
individual reliabilities: (1) The simplest case is identification by
position-coincidence between two wavebands (Sec.~\ref{pwpcid}),
associating NVSS radio sources with RBSC X-ray sources, for example.
For each candidate we derive the probability that it is the correct
identification (the reliability of that identification).  (2) If the
positions are not sufficiently accurate to guarantee reliable
identifications by themselves, they may be supplemented by additional
data (Sec.~\ref{pwc}).  To continue our example, RBSC X-ray sources
have a flatter radio flux-density distribution than NVSS background
sources, so radio flux densities affect identification reliabilities.
The identification reliabilities derived directly in this section are
similar to those obtained via the ``likelihood ratio'' method by
Sutherland \& Saunders (1992).  (3) Even with the aid of additional
data, the RBSC position errors are too large for making reliable X-ray
identifications with faint optical objects.  We show how accurate
positions of NVSS radio sources associated with RBSC sources can be
used to select the correct optical counterparts.  The reliabilities of
such multiwavelength linked position-coincidence identifications are
derived in (Sec.~\ref{lpccid}). (4) Finally, linked
cross-identifications can themselves be strengthened by applying
additional constraints (e.g., optical magnitudes), and their
reliabilities are obtained in Section~\ref{lcidwc}.

\subsection{Direct Position-coincidence Identifications}\label{pwpcid}

All identification programs begin with a set of identification
candidates in some search area surrounding each source to be
identified.  This area should encompass all plausible candidates, but
its exact size and shape are not critical.  A larger than necessary
search area increases the number of candidates which must be evaluated 
but does not significantly degrade identification reliability.  We
used search circles of 180$\arcsec$ in radius (Figure~\ref{fig:diagram}) to identify
RBSC sources having rms positional uncertainties $\sigma_{\rm x}
\approx 10''-20''$.  The search radius is much larger than the
$3\sigma$ X-ray error circle to permit identification 
with extended and asymmetric radio sources.  Each search area contains
some numbers $m \geq 0$ of NVSS radio sources and $l \geq 0$ of
optical objects from the USNO catalog.

\placefigure{fig:diagram}

We consider radio identifications of X-ray sources using only the
X-ray and radio positions.  Let $P(R)$ ($R =1,\dots,m$) represent the
probability that the $R^{th}$ candidate is the correct radio
identification of an X-ray source and $P(0)$ the probability that none
is.  Since there is only a negligible chance that the radio
identification is exterior to the search area, $P(0)$ is just the
probability that the actual identification is too faint to be
recognized as a candidate.  For example, we cannot identify radio
counterparts fainter than the 2.5 mJy beam$^{-1}$ NVSS catalog limit.
Such an identification is often called an ``empty field.''  If we make
the {\it astronomical} assumption that not more than one of the $m$
candidates is the correct identification, then the sum of these
mutually exclusive probabilities is unity:
\begin{equation}\label{pnormeq}
\sum_{R = 0}^m P(R)= 1~.
\end{equation}

The $R^{th}$ candidate is the correct identification if (1) there exists
a detectable identification, (2) that identification lies in the
infinitesimal area element $dA$ containing the position of the $R^{th}$
candidate, and (3) the ($m-1$) remaining candidates are unrelated
sources which happen to lie in the areas $dA$ surrounding their
positions (see Figure~\ref{fig:diagram}). The probability that all
three independent events occur is the product of their individual
probabilities, which we now evaluate.

(1) The {\it a priori} probability that there is a detectable radio
identification is equal to the initially unknown fraction $f_{\rm r}$
of X-ray sources in the sample that actually have detectable radio
identifications.  We estimated $f_{\rm r}$ by guessing an initial
value, making trial identifications for the whole source sample,
replacing the initial value by the observed value, and iterating. If
this procedure fails to converge rapidly, the resulting
identifications are probably too unreliable to be useful.  To estimate
the observed value of $f_{\rm r}$ we summed the computed reliabilities
of all the RBSC-NVSS identifications and divided that sum by the total
number of X-ray sources.  Since only a small percentage of bright galactic
X-ray stars are radio sources, we calculated separate $f_{\rm r}$
values for fields containing optically bright ($m < 12$) stars and for
all other fields.

(2) The probability of finding the correct identification in the
infinitesimal area $dA$ offset by the angle $\phi_{\rm xr}(R)$ between
the position of the X-ray source and the $R^{th}$ candidate is
$p[\phi_{\rm xr} (R)] dA$, where $p[\phi_{\rm xr}(R)]$ is just the
normalized error distribution of the measured X-ray/radio offsets.
These errors include both the radio and X-ray measurement errors and
may be augmented by a contribution allowing for possible astronomical
offsets of extended sources.  For example, the centroid of a head-tail
radio source will not coincide with its parent galaxy.  

(3) If the $R^{th}$ candidate is the correct identification, then the
remaining $(m-1)$ candidates must be unrelated sources lying in areas
$dA$ containing their positions.  The probability of finding each
unrelated source is $\rho_{\rm r}dA$, where $\rho_{\rm r}$ is the mean
sky density of unrelated radio candidates.  For NVSS candidates,
\begin{equation}\label{rhoeq}
\rho_{\rm r} = \int_{S_{\rm min}}^\infty n(S)dS ~,
\end{equation}
where $n(S)$ is the differential source count at 1.4 GHz and $S_{\rm
min} = 2.5$ mJy is the minimum flux density of the candidates.
Possible clustering of candidates around the true identification could
be addressed by increasing $\rho_{\rm r}$ from its global to local
value.  If there is no detectable identification, all $m$ candidates
must be unrelated sources.  

The resulting set of $(m+1)$ proportionalities
\begin{mathletters}\label{preleq}
\begin{eqnarray}
P(R) \propto f_{\rm r} p[\phi_{\rm xr}(R)] dA (\rho_{\rm r} dA)^{m-1}
     {\rm\qquad if}~R>0
\end{eqnarray}
\begin{eqnarray}
P(0) \propto (1 - f_{\rm r}) (\rho_{\rm r} dA)^m 
\end{eqnarray}
\end{mathletters}
normalized by Equation~\ref{pnormeq} specify the $(m+1)$ identification
reliabilities:
\begin{mathletters}\label{peq}
\begin{eqnarray}
P(R) = c f_{\rm r}
{p[\phi_{\rm xr}(R)] \over \rho_{\rm r}} 
     {\rm\qquad if}~R>0
\end{eqnarray}
\begin{eqnarray}
P(0) = c (1 - f_{\rm r})~,
\end{eqnarray}
where 
\begin{eqnarray}
c^{-1} = {f_{\rm r} \over \rho_{\rm r} }
\sum_{j = 1}^m  p[\phi_{\rm xr}(j)]
+ (1 - f_{\rm r})~.
\end{eqnarray}
\end{mathletters}
Equation~\ref{peq} gives the probability $P(R)$ that the $R^{th}$ of
$m$ candidates is the correct identification 
and the probability $P(0)$ that none is.  These probabilities depend
on the search area only through the number $m$ of candidates
contributing to the sum in Equation~\ref{peq}c.  Making the search
area ``too big'' by definition only adds candidates having negligible
$p(\phi_{\rm xr})$ and hence little effect in Equation~\ref{peq}c.

In many cases, the positional error distributions of both the source
and its identification candidates are nearly circular Gaussians.  [See
\citet{condon95} for the more complicated case of elliptical
Gaussian error distributions with arbitrary orientations.]  If the
X-ray source and radio candidate positions have rms uncertainties
$\sigma_{\rm x}$ and $\sigma_{\rm r}$ in each coordinate, then
\begin{equation}
p(\phi_{\rm xr}) = {1 \over 2\pi\sigma^2} \exp \biggl( -
{\phi_{\rm xr}^2 \over 2 \sigma^2} \biggr)~,
\end{equation}
where $\sigma^2 \equiv \sigma_{\rm x}^2 + \sigma_{\rm r}^2$.  The
variance of $\phi_{\rm xr}$ is $\langle \phi_{\rm xr}^2
\rangle = 2\sigma^2$ so a typical identification has $p(\phi_{\rm xr})
\sim (2 \pi \sigma^2 e)^{-1}$ and is highly reliable if both
\begin{mathletters}\label{posacceq}
\begin{eqnarray}
\sigma^2 \ll { f_{\rm r} \over 2 \pi e \rho_{\rm r} (1 - f_{\rm r})}
\end{eqnarray}
and
\begin{eqnarray}
\sigma^2 \ll {1 \over 2 \pi e \rho_{\rm r}}
\end{eqnarray}
\end{mathletters}
Equation~\ref{posacceq}a ensures that true empty fields are not
misidentified with background sources, a danger if the actual
identification rate is very low ($f_{\rm r} \ll 1$).
Equation~\ref{posacceq}b ensures that there is no difficulty choosing
the correct identification from multiple candidates.  For NVSS
identifications of extragalactic RBSC sources, $f_{\rm r} \approx
0.61$ and the sky density of sources stronger than the NVSS catalog
limit $S_{\rm min} \approx 2.5$~mJy is $\rho_{\rm r} \approx 1.76
\times 10^5$~sr$^{-1}$, leading to the fairly weak requirement $\sigma
\ll 10^2$ arcsec which is easily satisfied by the RBSC positional
uncertainties.  The sky density of optical objects in the USNO catalog
is two orders of magnitude higher, and the RBSC positional
uncertainties are too large to satisfy the requirement $\sigma \ll 10$
arcsec for making optical identifications by position coincidence
alone, even though most RBSC sources do have optical counterparts in
the USNO catalog.

\subsection{Pairwise Identifications With Additional Constraints}\label{pwc}

Uncertain identifications based on positional coincidence alone may be
strengthened or rejected by non-positional data. For example, the
flux-density distribution of radio sources identified with
extragalactic RBSC sources peaks well above the NVSS sensitivity
limit. Figure~\ref{fig:radsrc} shows the logarithm of the ratio of the
probability $p[S\vert x]$ that a RBSC-NVSS source has flux density
$S$ to the probability $p[S\vert \overline{x}]$ that an unrelated NVSS
sources has flux density $S$. Most
unrelated radio sources are fainter than the correct identifications
(indicated in Figure~\ref{fig:radsrc} by the positive logarithm of the
ratio at all but the faintest radio flux levels), so the stronger of
two radio candidates with similar $p(\phi_{\rm xr})$ is the more
likely identification. The reliabilities of such identifications can
be calculated through the use of ``likelihood ratios''
\citep[see][]{sutherland92}; here we obtain the identification
reliabilities directly by extending the derivation of
Equation~\ref{peq}.

\placefigure{fig:radsrc}

Let $p[\phi_{\rm xr}(R), S(R)] dA dS$ be the probability that the
$R^{th}$ candidate is the identification lying in the area $dA$
surrounding its observed position and in the flux-density range $dS$
containing its observed flux density $S(R)$.  The multiplicative law
of probabilities states that this probability is the product of
$p[\phi_{\rm xr}(R)]dA$ and $p[S(R) \vert x] dS$, where $p(S \vert x)$
is the normalized ($\int_{S_{\rm min}} ^\infty p(S \vert x) dS = 1$)
flux-density distribution of radio sources identified with X-ray
sources.  Like $f_{\rm r}$, $p(S \vert x)$ is best estimated from the
actual identification data by iteration.  The probability that the
$R^{th}$ candidate is an unrelated (X-ray quiet) radio source with flux
density $S$ lying in area $dA$ is $p [S(R) \vert \overline{x}]
\rho_{\rm r} dA dS$, where $p (S \vert \overline{x})$ is the
flux-density distribution of all (background) NVSS sources.  Thus
Equations~\ref{preleq} and \ref{peq} can be replaced by:
\begin{mathletters}\label{genprop}
\begin{eqnarray}
P(R) \propto f_{\rm r} p[\phi_{\rm xr}(R)] 
{ p[S(R) \vert x] \over p[S(R) \vert \overline{x}]} dA dS
\biggl\{ \prod_{j =1}^m p[S(j) \vert \overline{x}] \biggr\} 
(\rho_{\rm r} dA dS)^{m-1}
  {\rm\qquad if}~R>0
\end{eqnarray}
\begin{eqnarray}
P(0) \propto (1 - f_{\rm r}) \biggl\{ \prod_{j = 1}^m 
p[S(j)] \vert \overline{x}] \biggr\} (\rho_{\rm r}dA dS)^m
\end{eqnarray}
\end{mathletters}
and
\begin{mathletters}\label{genid}
\begin{eqnarray}
P(R) = c  f_{\rm r} 
{ p[\phi_{\rm xr}(R)] \over \rho_{\rm r} }
 {p[S(R) \vert x] \over p[S(R) \vert 
\overline{x}] }
     {\rm\qquad if}~R>0
\end{eqnarray}
\begin{eqnarray}
P(0) = c (1 - f_{\rm r})
\end{eqnarray}
where
\begin{eqnarray}
c^{-1} = {f_{\rm r} \over \rho_{\rm r} }
\sum_{j=1}^m  p[\phi_{\rm xr}(j)]
{p[S(j) \vert x] \over p[S(j) \vert 
\overline{x}] } + (1 - f_{\rm r})
\end{eqnarray}
\end{mathletters}
Equation~\ref{genid} gives the identification probabilities based on
candidate flux densities as well as positional coincidence.  It could
be further extended to include additional continuous (e.g., spectral
index) or discrete (e.g., morphological type) parameters which might
prove useful for distinguishing between correct identifications and
unrelated candidates.

\subsection{Linked Position-coincidence
Cross-identifications}\label{lpccid}

The positional uncertainties of most RBSC sources are too large to
yield reliable optical identifications with faint galaxies and quasars
directly.  However, the NVSS sources reliably identified with RBSC
sources have sufficiently accurate radio positions that nearly all can
be optically identified by position coincidence alone.  The
reliabilities of such linked X-ray/radio/optical position-coincidence
identifications are derived below.

Let $P(R,V)$ be the probability that the $R^{th}$ radio source and the
$V^{th}$ optically visible object are the correct identifications of an
X-ray source, where $R = 0,1,\dots, m$ and $V = 0,1,\dots,l$.  The
values $R=0$ and $V=0$ correspond to radio and optical ``empty
fields,'' respectively.  The probabilities of these $(m+1)(l+1)$
mutually exclusive possible outcomes must add up to one:
\begin{equation}\label{2dnormeq}
\sum_{R=0}^m \sum_{V=0}^l P(R,V) = 1~.
\end{equation}
Denote the fractions of X-ray sources in the sample having detectable
($R>0$) radio and optical ($V>0$) identifications by $f_r$ and $f_v$,
respectively.  Let $p[\phi_{\rm xr}(R)]$ and $p[(\phi_{\rm xv}(V)]$ be
the probability densities of the X-ray/radio and X-ray/optical
positional offsets $\phi$ if the $R^{th}$ radio source and the
$V^{th}$ optical object are the correct identifications.  The sky
densities of 
background radio and optical candidates are $\rho_{\rm r}$ and
$\rho_{\rm v}$, respectively.  Using the multiplication law $P(R,V) =
P(R) P(V \vert R)$ yields:
\begin{mathletters}\label{probpropeq}
\begin{eqnarray}
P(R,V) \propto f_{\rm r} p[\phi_{\rm xr}(R)]dA (\rho_{\rm r} dA)^{m-1}
     f_{\rm v} p[\phi_{\rm xv}(V)]dA (\rho_{\rm v} dA)^{l-1} 
     {\rm\qquad if}~R>0,V>0
\end{eqnarray}
\begin{eqnarray}
P(R,0) \propto f_{\rm r} p[\phi_{\rm xr}(R)]dA (\rho_{\rm r} dA)^{m-1}
     (1 - f_{\rm v}) (\rho_{\rm v} dA)^l
     {\rm\qquad if}~R>0
\end{eqnarray}
\begin{eqnarray}
P(0,V) \propto (1 - f_{\rm r}) (\rho_{\rm r} dA)^m
     f_{\rm v} p[\phi_{\rm xv} (V)] dA (\rho_{\rm v} dA)^{l-1}
     {\rm\qquad if}~V>0
\end{eqnarray}
\begin{eqnarray}
P(0,0) \propto (1 - f_{\rm r}) (\rho_{\rm r} dA)^m
     (1 - f_{\rm v}) (\rho_{\rm v} dA)^l
\end{eqnarray}
\end{mathletters}
The normalization Equation~\ref{2dnormeq} then implies:
\begin{mathletters}\label{2dprobeq}
\begin{eqnarray}
P(R,V) = c f_{\rm r} f_{\rm v} 
     {p[\phi_{\rm xr} (R)] \over \rho_{\rm r}}
     {p[\phi_{\rm xv} (V)] \over \rho_{\rm v}}
     {\rm\qquad if}~R>0,V>0
\end{eqnarray}
\begin{eqnarray}
P(R,0) = c f_{\rm r} (1 - f_{\rm v})
     {p[\phi_{\rm xr} (R)] \over \rho_{\rm r}}
     {\rm\qquad if}~R>0
\end{eqnarray}
\begin{eqnarray}
P(0,V) = c (1 - f_{\rm r}) f_{\rm v}
     {p[\phi_{\rm xv} (V)] \over \rho_{\rm v}}
     {\rm\qquad if}~V>0
\end{eqnarray}
\begin{eqnarray}
P(0,0) = c (1 - f_{\rm r}) (1 - f_{\rm v})~,
\end{eqnarray}
where
\begin{eqnarray}\nonumber
c^{-1} = { f_{\rm r} f_{\rm v} \over \rho_{\rm r} \rho_{\rm v} }
\sum_{j=1}^m
     p[\phi_{\rm xr} (j)]
     \sum_{i=1}^l
     p[\phi_{\rm xv} (i)] 
\end{eqnarray}
\begin{eqnarray}
     + {f_{\rm r} (1 - f_{\rm v}) \over \rho_{\rm r}}
     \sum_{j=1}^m  p[\phi_{\rm xr} (j)]
     + {(1 - f_{\rm r}) f_{\rm v} \over \rho_{\rm v}}
     \sum_{i=1}^l p[\phi_{\rm xv} (i)]
     + (1 - f_{\rm r}) (1 - f_{\rm v})~.
\end{eqnarray}
\end{mathletters}

Unfortunately, applying Equation~\ref{2dprobeq} does not yield good
optical identifications of RBSC X-ray sources because both the X-ray
positional uncertainties and the mean density $\rho_{\rm v}$ of
optical candidates are large.  Reliable radio identifications of X-ray
sources can be made because the density $\rho_{\rm r}$ of radio
candidates is much smaller, and reliable optical
identifications of NVSS radio sources are possible because the NVSS
positions are more accurate.  If the {\it astronomical} assumption
is made that the optical identifications of these radio
identifications are also the optical identifications of the
corresponding X-ray sources, then

\begin{mathletters}\label{linkpropeq}
\begin{eqnarray}\nonumber
P(R,V) \propto f_{\rm r} p[\phi_{\rm xr}(R)]dA (\rho_{\rm r} dA)^{m-1}
     f_{\rm v} p[\phi_{\rm xv}(V)]dA (\rho_{\rm v} dA)^{l-1}
     p[\phi_{\rm rv}(R,V)]dA (\rho_{\rm v} dA)^{l-1} 
\end{eqnarray}
\begin{eqnarray}
\hphantom{xxxxxxxxxxxxxxxxxxxxxxxxxxxxxxx}     {\rm\qquad if}~R>0,V>0
\end{eqnarray}
\begin{eqnarray}
P(R,0) \propto f_{\rm r} p[\phi_{\rm xr}(R)]dA (\rho_{\rm r} dA)^{m-1}
     (1 - f_{\rm v}) (\rho_{\rm v} dA)^l (\rho_{\rm v} dA)^l
     {\rm\qquad if}~R>0
\end{eqnarray}
\begin{eqnarray}
P(0,V) \propto (1 - f_{\rm r}) (\rho_{\rm r} dA)^m
     f_{\rm v} p[\phi_{\rm xv} (V)] dA (\rho_{\rm v} dA)^{l-1}
     (\rho_{\rm v} dA)^l
     {\rm\qquad if}~V>0
\end{eqnarray}
\begin{eqnarray}
P(0,0) \propto (1 - f_{\rm r}) (\rho_{\rm r} dA)^m
     (1 - f_{\rm v}) (\rho_{\rm v} dA)^l (\rho_{\rm v} dA)^l ~,
\end{eqnarray}
\end{mathletters}
where $p[\phi_{\rm rv} (R,V)]$ is the
probability distribution of offsets between the $R^{th}$ radio source and
the $V^{th}$ optical object.
The normalization Equation~\ref{2dnormeq} implies
\begin{mathletters}\label{linkprobeq}
\begin{eqnarray}
P(R,V) = c f_{\rm r} f_{\rm v} 
     {p[\phi_{\rm xr} (R)] \over \rho_{\rm r}}
     {p[\phi_{\rm xv} (V)] \over \rho_{\rm v}}
     {p[\phi_{\rm rv} (R,V)] \over \rho_{\rm v}}
     {\rm\qquad if}~R>0,V>0
\end{eqnarray}
\begin{eqnarray}
P(R,0) = c f_{\rm r} (1 - f_{\rm v})
     {p[\phi_{\rm xr} (R)] \over \rho_{\rm r}}
     {\rm\qquad if}~R>0
\end{eqnarray}
\begin{eqnarray}
P(0,V) = c (1 - f_{\rm r}) f_{\rm v}
     {p[\phi_{\rm xv} (V)] \over \rho_{\rm v}}
     {\rm\qquad if}~V>0
\end{eqnarray}
\begin{eqnarray}
P(0,0) = c (1 - f_{\rm r}) (1 - f_{\rm v})~,
\end{eqnarray}
where
\begin{eqnarray}\nonumber
c^{-1} = { f_{\rm r} f_{\rm v} \over \rho_{\rm r} \rho_{\rm v}^2 }
\sum_{j=1}^m
     p[\phi_{\rm xr} (j)]
     \sum_{i=1}^l
     p[\phi_{\rm xv} (i)] 
     p[\phi_{\rm rv} (j,i)]
\end{eqnarray}
\begin{eqnarray}
     + {f_{\rm r} (1 - f_{\rm v}) \over \rho_{\rm r}}
     \sum_{j=1}^m  p[\phi_{\rm xr} (j)]
     + {(1 - f_{\rm r}) f_{\rm v} \over \rho_{\rm v}}
     \sum_{i=1}^l p[\phi_{\rm xv} (i)]
     + (1 - f_{\rm r}) (1 - f_{\rm v})~.
\end{eqnarray}
\end{mathletters}
Equation~\ref{linkprobeq} specifies the reliabilities of
linked X-ray/radio/optical identifications made on the basis
of position-coincidence alone.

\subsection{Linked 
Cross-identifications With Additional Constraints}\label{lcidwc}

\placefigure{fig:optsrc}

Finally, differences between the magnitude distributions of the
optical counterparts to X-ray sources and background optical objects
can be used to improve the reliability of the optical identifications.
Figure~\ref{fig:optsrc} gives the logarithm of the ratio of the probability
$p[\mu\vert x]$ that the USNO identification of a RBSC-NVSS source
has magnitude $\mu$ to the probability $p[\mu\vert \overline{x}]$ that
an unrelated USNO object has magnitude $\mu$ as a function of B
magnitude, and shows, in contrast to the radio, that there is very
little difference between the two populations for B $> 13$, but the
ratio does become 
large for objects brighter than this limit. Let $\mu(V)$ denote the
magnitude of the $V^{th}$ optical candidate. Let $p(\mu \vert x)$ and
$p(\mu \vert 
\overline{x})$ be the normalized magnitude distributions of optical
objects in the USNO catalog which are X-ray sources and unrelated
optical objects, respectively. Let $p(S \vert x)$ be the normalized
flux-density distribution of radio-detected X-ray sources and $p(S
\vert \overline{x})$ be the the normalized flux-density distribution
of unrelated radio sources stronger than the NVSS limit, 2.5 mJy.
Then
\begin{mathletters}\label{2dgenprop}
\begin{eqnarray}\nonumber
P(R,V) \propto 
f_{\rm r} p[\phi_{\rm xr} (R)] 
{ p[S(R) \vert x] \over  p[S(R) \vert \overline{x}] } dA dS
 \biggl\{ \prod_{j=1}^m p[S(j) \vert \overline{x}] \biggr\}
 (\rho_{\rm r} dA dS)^{m-1}
\end{eqnarray}
\begin{eqnarray}\nonumber
\times f_{\rm v} p[\phi_{\rm xv} (V)] { p[\mu (V) \vert x] \over
p[\mu(V) \vert \overline{x}] } dA d\mu \biggl\{
\prod_{i=1}^l p[\mu (i) \vert \overline{x}] \biggr\}
(\rho_{\rm v} dA d\mu)^{l-1}
\end{eqnarray}
\begin{eqnarray}
\times p[\phi_{\rm rv}(R,V)] dA (\rho_{\rm v} dA)^{l - 1} \qquad
 {\rm\qquad if}~R>0,V>0
\end{eqnarray}
\begin{eqnarray}\nonumber
P(R,0) \propto
f_{\rm r} p[\phi_{\rm xr} (R)] 
{ p[S(R) \vert x] \over p[S(R) \vert \overline{x}] } dA dS
\biggl\{ \prod_{j=1}^m p[S(j) \vert \overline{x}] \biggr\}
 (\rho_{\rm r} dA dS)^{m-1}
\end{eqnarray}
\begin{eqnarray}
\times (1-f_{\rm v})
\biggl\{ \prod_{i=1}^l p[\mu (i) \vert \overline{x}] \biggr\} 
(\rho_{\rm v} dA d\mu)^l (\rho_{\rm_v} dA)^l
     {\rm\qquad if}~R>0
\end{eqnarray}
\begin{eqnarray}\nonumber
P(0,V) \propto
(1 - f_{\rm r})
\biggl\{ \prod_{j=1}^m p[S(j) \vert \overline{x}] \biggr\} 
(\rho_{\rm r} dA dS)^m
\end{eqnarray}
\begin{eqnarray}
\times f_{\rm v} 
 p[\phi_{\rm xv} (V)] 
{ p[\mu(V) \vert x]\over p [\mu(V) \vert \overline{x}] }
  dA d\mu
\biggl\{ \prod_{i=1}^l p[\mu (i) \vert \overline{x}] \biggr\} 
(\rho_{\rm v} dA d\mu)^{l-1} (\rho_{\rm_v} dA)^l
     {\rm\qquad if}~V>0
\end{eqnarray}
\begin{eqnarray}\nonumber
P(0,0) \propto
(1 - f_{\rm r})
\biggl\{ \prod_{j=1}^m p[S(j) \vert \overline{x}] \biggr\} 
(\rho_{\rm r} dA dS)^m
\end{eqnarray}
\begin{eqnarray}
\times (1 - f_{\rm v})
\biggl\{ \prod_{i=1}^l p[\mu (i) \vert \overline{x}] \biggr\} 
(\rho_{\rm v} dA d\mu)^l  (\rho_{\rm_v} dA)^l~.
\end{eqnarray}
\end{mathletters}
The normalized probabilities are:
\begin{mathletters}\label{2dgenid}
\begin{eqnarray}\nonumber
P(R,V) = c
f_{\rm r} f_{\rm v} {p[\phi_{\rm xr} (R)] \over \rho_{\rm r}} 
{ p[S(R) \vert x] \over p[S(R) \vert \overline{x}]}
{ p[\phi_{\rm xv} (V)] \over \rho_{\rm v}} 
{p[\mu (V) \vert x] \over p [\mu(V) \vert \overline{x}] }
{ p[\phi_{\rm rv}(R,V)] \over \rho_{\rm v} }
\end{eqnarray}
\begin{eqnarray}
\hphantom{xxxxxxxxxxxxxxxxxxxxxxxxxxxxxxx}     {\rm\qquad if}~R>0,V>0
\end{eqnarray}
\begin{eqnarray}
P(R,0) = c
f_{\rm r} (1-f_{\rm v})  {p[\phi_{\rm xr} (R)] \over \rho_{\rm r}}
{ p[S(R) \vert x] \over p[S(R) \vert \overline{x}]} 
     {\rm\qquad if}~R>0
\end{eqnarray}
\begin{eqnarray}
P(0,V) = c
(1 - f_{\rm r})
f_{\rm v} 
{p[\phi_{\rm xv} (V)] \over \rho_{\rm v} } 
{p[\mu(V) \vert x] \over p [\mu(V) \vert  \overline{x}] }
     {\rm\qquad if}~V>0
\end{eqnarray}
\begin{eqnarray}
P(0,0) = c
(1 - f_{\rm r})
(1 - f_{\rm v})
~,
\end{eqnarray}
where
\begin{eqnarray}\nonumber
c^{-1} = 
f_{\rm r} f_{\rm v} 
 \sum_{j=1}^m
{p[\phi_{\rm xr} (j)] \over \rho_{\rm r}}
{ p[S(j) \vert x] \over p[S(j) \vert \overline{x}]}
 \sum_{i=1}^l  { p[\phi_{\rm xv} (i)] \over \rho_{\rm v}} 
 {p[\mu (i) \vert x] \over p [\mu(i) \vert \overline{x}] } 
 {p[\phi_{\rm rv}(j,i)] \over \rho_{\rm v} }
\end{eqnarray}
\begin{eqnarray}\nonumber
+ f_{\rm r} (1-f_{\rm v}) 
\sum_{j=1}^m
{p[\phi_{\rm xr} (j)] \over \rho_{\rm r}} 
{ p[S(j) \vert x] \over p[S(j) \vert \overline{x}]}
+(1 - f_{\rm r}) f_{\rm v} 
\sum_{i=1}^l
{p[\phi_{\rm xv} (i)] \over \rho_{\rm v} } 
{p[\mu(i) \vert x] \over p [\mu(i) \vert \overline{x}] }
\end{eqnarray}
\begin{eqnarray}
+ (1 - f_{\rm r}) (1 - f_{\rm v})
~.
\end{eqnarray}
\end{mathletters}
It is with Equation~\ref{2dgenid} that we evaluated the reliabilities
of the RBSC identification candidates found in Section~\ref{results}.

The final identification process proceeded as follows. We 
started with a list of X-ray positions and errors from the RBSC. We
then searched for corresponding NVSS radio and USNO optical sources
within a $3\arcmin$ radius ($R =0,\dots,m$ and $V =0,\dots,l$
respectively). This particular search size was used because it was
much larger than the 3$\sigma$ X-ray error circle and large enough to
contain all but the most extended radio and optical sources. The
positions of these sources (X-ray, radio, and optical), their
positional errors, and their fluxes and magnitudes were combined with
the {\it a priori} probabilities of detecting radio and optical
counterparts, $f_{\rm r}$ and $f_{\rm v}$, the RBSC radio flux and
optical magnitude distributions using Equation~\ref{2dgenid} to
estimate the probabilities, $P(R,V)$. We estimated $f_{\rm r}$ and
$f_{\rm v}$ initially , and then iteratively replaced these values
with the actual identification rates until they converged to the
values of 0.61 and 0.99. The RBSC flux and magnitude distributions
were again found iteratively. The radio flux distribution of RBSC
sources was compared to the distribution of background NVSS objects
and the RBSC optical magnitude distribution was compared to the
position dependent optical background distribution given by
\citet{bahcall80}.

\subsection{Additional Parameters and Possible Concerns}\label{lcidwc_ap}

Many astronomical objects have extended X-ray, radio, or optical
structures which lead to offsets between wavebands beyond those
accounted for by measurement errors. 

To allow for an offset between the centroid and central component of a
radio source smaller than the NVSS beam, we added one tenth of the
deconvolved NVSS size or upper limit in quadrature to the rms
uncertainty of the NVSS centroid position. Further, if this RBSC-NVSS
matched field happened to lie within the area covered by the VLA FIRST
Survey \citep[][$5\arcsec$ resolution, 1 mJy flux limit]{white97}, we
then evaluated Equation~\ref{2dgenid}, and replaced the NVSS sources
with the FIRST sources if the reliability was better.

An attempt was also made to identify sources resolved into two or more
NVSS components, such as double-lobed or head-tail radio galaxies. If
two radio components in the search area were within $3\arcmin$ of one
another and had a flux ratio $<3$, we initiated a test for a double
source. This consisted of adding an artificial ``source'' with the
combined flux of the two components and located at their radio
centroid (this gave much better results than the {\it optimally}
weighted centroid used in Windhorst et al. 1984) to the candidate
list. We then re-evaluated Equation~\ref{2dgenid}, keeping this new
``source'' if the reliability was greater than 0.8 or reverting back
to the original reliability estimation otherwise. The rms major- and
minor-axis position errors for these artificial sources were taken to
be $\slantfrac{1}{10}$ and $\slantfrac{1}{30}$ the size of the
component separation. We also found a few true double sources with
flux ratios $>3$ which were evaluated in the same manner.

We have not corrected for extended RBSC sources.  This has the effect
of selecting against large scale X-ray emission not coincident with
optical and radio sources, such as clusters of galaxies. 

Since the USNO catalog does not list optical angular sizes, we used
the AIPS Gaussian fitting task IMFIT to measure the optical diameters
of of objects larger than $0\farcm5$.  An additional error of
$\slantfrac{1}{15}$ of the major axis was added in quadrature with the
rms optical positional uncertainty (which we took to be $0\farcs5$).

The USNO $O$ plate magnitudes are converted into Johnson
B magnitudes using $B = O - 0.119(O-E)$ \citep{evans88}. 
We found that these USNO magnitudes are systematically too bright for
galaxies with B $\la 16$ by as much as three
magnitudes in both R and B as shown by a comparison of USNO magnitudes 
with B magnitudes from the NASA/IPAC Extragalactic Database
\citep{ned99} (Figure~\ref{fig:mB_vs_mBned}). There is better
agreement (the dispersion is $\sim 2$ mag) between USNO and NED
magnitudes for galaxies with B $\ga 
16$. To avoid overestimating the identification reliabilities of
objects brighter than B $\approx 16$, we corrected the USNO blue
magnitudes using the polynominal fit
\begin{eqnarray}\nonumber
\rm{B}_{\rm{NED}} = -0.066 + 1.172\rm{B}_{\rm{USNO}} + 0.120(\rm{B}_{\rm{USNO}})^2 -
0.014(\rm{B}_{\rm{USNO}})^3 + 0.000376(\rm{B}_{\rm{USNO}})^4
\end{eqnarray}

\noindent indicated by the solid curve in
Figure~\ref{fig:mB_vs_mBned}.  The remaining scatter ($\pm$ 2 mag) is
not large enough to impact the reliability of our identifications
since the ratio $p(\mu \vert x) / p(\mu \vert \overline{x})$ does not
change by more than a factor of a few over the affected magnitude
range (Fig.~\ref{fig:optsrc}). 

One problem associated with catalogued flux densities and magnitudes
is Malmquist bias. Given some error in the flux and magnitude
determination and the steep slope of the counts, the true brightness
of an object will be overestimated. While this is a potential pitfall
for faint flux and magnitude counts, it should not significantly
affect the reliabilities calculated here for several reasons: 1) the
flux and magnitude estimates are normalized by the same Malmquist
biased background, 2) for the optical sources, the flat magnitude
distribution (Fig.~\ref{fig:optsrc}) carries little weight in the
reliability determination anyway, and 3) the slope of the optical (and
to a lesser extent the radio) RBSC source counts turns over well in
advance of the survey limit, such that there are very few faint
sources where the bias would be strongest.

\placefigure{fig:mB_vs_mBned}

A further complication is the high sky density of bright foreground
stars at low Galactic latitudes.  The likelihood that an optical star
is an X-ray emitter is quite large (one third of RBSC sources are
stellar in origin; \citep{bade98}), while the likelihood that the same
optical star is also a radio emitter is quite small ($\sim 0.1-2\%$,
see \citet{condon97}). Thus a bright X-ray star near a background
radio source may result in the identification of the X-ray star with
that radio source. Because the density of background optical sources
increases more rapidly toward fainter magnitudes than the density of
RBSC-NVSS optical identifications does, a bright optical star
will be assigned a greater identification reliability than a faint
galaxy, even if the star is not quite as close to the
radio position as the galaxy is.  Weeding out X-ray objects which are
likely to be stars will then decrease the chances of calling an
interloper the correct identification. Thus we classified all objects brighter
than B $< 13$ mag on the basis of appearance on the Digitized Sky
Survey \citep[][DSS]{lasker90} as stars (with diffraction spikes,
saturated point source, no diffuse halo) or galaxies.  X-ray fields
containing one or more bright stars but no bright galaxies within
$3\sigma$ of the X-ray position were considered stellar. In these
cases, the value of $f_r$ was forced to be $0.02$ for stellar fields
rather than $0.61$, the extragalactic identification fraction.  We may
have missed a few radio stars with high proper motions because the
USNO2 and radio positions were measured at different epochs.

Finally, we used the locally measured, rather than global, surface
density $\rho_{\rm v}$ of background objects because many RBSC sources
reside in optically overdense regions (clusters of galaxies).

\section{Identification Results}\label{results}

For the parent sample, we selected the 5441 RBSC sources
above the 0.1 counts s$^{-1}$ completeness limit with Galactic
latitudes $|b| > 15^\circ$ (to reduce the number of objects which
might confuse optical identification and minimize extinction at both
optical and X-ray wavelengths) and J2000 $\delta > -40^\circ$, the
NVSS declination limit. Of these, 1773 sources are readily
identified as stars, in agreement with \citet{bade98}. We attempted to
make {\it linked} X-ray/radio/optical 
cross-identifications using X-ray/radio and radio/optical positional
coincidence supplemented by radio flux densities and optical
magnitudes, as described in Section~\ref{lcidwc}.

We visually confirmed each match by overlaying the contours of the
NVSS image and the RBSC error box on the DSS (see
Figure~\ref{fig:charts}). In addition, the IRAS Faint Source
Catalogue \citep[][FSC]{moshir92} has been matched against the RBSC
\citep[see][]{condon98b}, so we have plotted the FSC 3$\sigma$
error ellipses to aid identification and differentiate between
physical emission processes.
The nearest candidate usually has the highest calculated reliability
(Equation~\ref{2dgenid}), although the radio emission for some sources
is extended and more care had to be taken in locating an optical
counterpart. We use the term {\it association} to describe extended
X-ray/radio sources which appear related but may not be spatially
coincident. For instance, a cluster of galaxies may contain one or
more radio galaxies and hot intercluster X-ray gas.
We retained 19 identifications with low calculated
reliabilities on the basis of other information, such as
detection in a {\it ROSAT} pointed observation or no obvious optical
counterpart within the error circle. All
deviations from the standard reliability estimate are noted in Section 
\ref{idnotes}. The reliabilities of these sources are given as '$-$'. 

\placefigure{fig:charts}

Some RBSC-NVSS sources had no USNO sources listed near their NVSS
positions, but objects were clearly visible on the DSS (i.e the red
plate). Most were either very bright, extended galaxies or very faint
objects. USNO sources were identified based on a positional
coincidence of within $2\farcs.0$ on the red and blue plates, so we
might expect that some bright galaxies will have red and blue emission
peaks separated by more than $2\farcs.0$ and thus be rejected. For
these we used positions and magnitudes from NED. Furthermore, the
faint objects we find must only appear on the red plate (the lack of
a blue plate counterpart presumably excluded them from the USNO
catalog). We used the AIPS task IMFIT to measure their positions.
Faint object magnitudes were estimated by comparison with nearby USNO
sources.

These additional sources were added to the USNO list of optical
candidates located within the search area and evaluated as part of the
standard reliability calculation (see Equation~\ref{2dgenid}). This
additional step introduces a slight bias in the reliabilities of these
new faint sources, because we have only added the faint sources in
close proximity to the radio/X-ray positions, and not those from the
rest of the $3\arcmin$ radius field. We can test the statistical error
of this effect by simply doubling the number of local background
objects; this is akin to probing roughly 2 magnitudes fainter than the
USNO. While the change will depend on the exact field, we found the
reliability with this background enhancement typically differed by
only 1\% (and at most 5\%).

The Galactic absorption corrected X-ray fluxes come from the RBSC
correlation catalogue \citep{voges99}, in which the RBSC count
rate was converted to flux assuming a power law photon energy
distribution of the form E$^{-\Gamma_{x} +1}$, common to AGN and
clusters of galaxies. An average photon index $\langle \Gamma_{\rm x}
\rangle = 2.3$ was used, typical of extragalactic 
objects (Hasinger, Tr\"{u}mper \& Schmidt 1991) over the {\it ROSAT}
energy range. Corrections for Galactic absorption were based on hydrogen
column densities  N$_{\rm H}$ obtained from \citet{dickey90}. 

The error of the unabsorbed X-ray flux is about 30\% taking into
account errors in photon statistics and variable Galactic absorption.
Flux estimates break down for cases when the actual 
$\Gamma_{\rm x}$ deviates significantly from the mean value, there is
additional intrinsic absorption, or 
when a single power law model inadequately describes the spectral
shape (see \citet{brinkmann95} for a more detailed discussion of
this problem).

\placefigure{fig:skydist}

A total of 1556 RBSC-NVSS sources were identified, on the criterion
that the sum total of all radio identification reliabilities be
greater than 0.50. The vast majority exceeded this cut-off limit a
large margin. The distribution of RBSC-NVSS sources on the sky is fairly  
uniform (Figure~\ref{fig:skydist}). Roughly
one half of the extragalactic RBSC sources did not have radio
counterparts. Only 1 RBSC-NVSS source has no detectable optical
counterpart (not surprising given the large RBSC error circles and the
density of optical sources at the limiting magnitude of the POSS).
Table~\ref{stars} lists the 44 possible galactic stars 
so identified because they have $B <$ 13 mag, diffraction spikes,
faint irregular radio contours or a spectral type from SIMBAD. Many
have been independently verified with high resolution VLA images to be
radio stars by \citet{condon97}. Table~\ref{stars} is arranged in the
following manner:

\small
\noindent {\bf Column 1}. {\it ROSAT} BSC name, format 1RXS
JHHMMSS.S+DDMMSS.
\\ \noindent {\bf Column 2}. SIMBAD Common Name, if available.
\\ \noindent {\bf Column 3}. {\it ROSAT} RBSC X-ray Flux.
\\ \noindent {\bf Column 4}. NVSS position, unless otherwise noted.
\\ \noindent {\bf Column 5}. NVSS 1.4 GHz Flux density.
\\ \noindent {\bf Column 6}. Identification Probability P(R,V)
(Equation~\ref{2dgenid}). Identifications with a '$-$' have been
accepted despite having a low probability.
\\ \noindent {\bf Column 7}. SIMBAD object type, if available. Abbreviations from SIMBAD as follows:
\begin{center}
\begin{tabular}{lcl}
* & = & Star \\
* Al & = & Eclipsing binary of Algol type \\
* bL & = & Eclipsing binary of Beta Lyr type \\
* Cl & = & Star or globular cluster \\
* DQ & = & Cataclysmic variable of DQ Her type \\
* Fl & = & Flare star \\
* Mi & = & Variable star of Mira Cet type \\
* Pu & = & Pulsating variable star \\
* Ro & = & Rotationally variable star \\
* RS & = & Star of RS CVn type \\
* sr & = & Semi-regular pulsating star \\
* TT & = & T Tau-type star \\
* V  & = & Variable star \\
PN & = & Planetary nebula
\end{tabular}
\end{center}
\noindent {\bf Column 8}. SIMBAD B magnitude, if available; otherwise
USNO magnitude enclosed in parentheses.
\\ \noindent {\bf Column 9}. SIMBAD spectral type, if available.
\normalsize

\placefigure{fig:reliability}

The remaining 1512 objects are extragalactic and listed in
Table~\ref{gals}. Figure~\ref{fig:reliability} shows the cumulative
fraction of identifications versus reliability. The vast majority
($\sim 70\%$) of identifications have reliabilities greater than 99\%.
Table~\ref{gals} is arranged in the following manner:

\small
\noindent {\bf Column 1}. {\it ROSAT} RBSC name, format 1RXS JHHMMSS.S+DDMMSS.
\\ \noindent {\bf Column 2}. NED Common Name, if available. 
\\ \noindent {\bf Column 3}. {\it ROSAT} BSC X-ray Flux.
\\ \noindent {\bf Column 4}. NVSS position, unless noted otherwise.
\\ \noindent {\bf Column 5}. NVSS 1.4 GHz Flux density.
\\ \noindent {\bf Column 6}. Identification Probability P(R,V)
(Equation~\ref{2dgenid}). Identifications with a '$-$' have been
accepted despite having a low probability.
\\ \noindent {\bf Column 7}. B magnitude from NED, if available;
otherwise from USNO as indicated by '()'. Identifications with a '$-$' 
denote empty fields.
\\ \noindent {\bf Column 8}. Redshift, if known. From NED unless noted
otherwise. Identifications with a '$-$' represent objects for which
spectra was taken but a redshift could not be determined. We
have obtained optical spectra with 
the KPNO 2.1 m telescope for many of the bright identifications which did not 
possess published redshifts. The detailed description of these spectra
will be given in a subsequent paper (Bauer, Thuan \& Condon 2000).
\\ \noindent {\bf Column 9}. NED optical morphology, if available.
\\ \noindent {\bf Column 10}. Spectral classification, if known. From NED unless noted otherwise in redshift column.
\begin{center}
\begin{tabular}{lcl}
AGN   & = & Active galactic nuclei, type unknown \\
Blazar & = & BL Lacertae type object with variable emission line spectra \\
BLLAC & = & BL Lacertae type object\\
BLRG  & = & Broad-line radio galaxy \\
cD    & = & Central dominant galaxy, Early-type stellar absorption continuum\\
Early & = & Early-type stellar absorption continuum \\
HII   & = & Starburst galaxy with narrow emission-line spectra similar to HII regions \\
HPQ   & = & High polarization QSO \\
LINER & = & Low ionization narrow emission-line region \\
LPQ   & = & Low polarization QSO \\
QSO   & = & Quasi-stellar object \\
Sy1-2 & = & Seyfert galaxy, emission-line spectra classified from type 1 to 2
\end{tabular}
\end{center}
\normalsize

There are a small number of cases having low reliabilities, but which 
may in fact be true identifications. They generally fall into two categories;
objects needing higher radio resolution and candidates that have
optical/radio matches farther than 3$\sigma$ from the X$-$ray
source which did not have a convincing optical match. Many of the latter
are likely to be associated with nearby clusters.

\placetable{stars}
\placetable{gals}

\subsection{Notes on Individual Identifications\label{idnotes}}
\scriptsize 
\noindent {\bf 1RXS J000312.3$-$355541}:  Member of cluster Abell 2717.
\\ \noindent {\bf 1RXS J001031.3$+$105832}:  Member of triple galaxy III Zw 002.
\\ \noindent {\bf 1RXS J001118.5$-$285126}:  Member of cluster Abell 2734.
Abell 2744. Diffuse radio emission; possible radio halo?
\\ \noindent {\bf 1RXS J001823.8$+$300357}:  Member of group WP 01.
\\ \noindent {\bf 1RXS J001906.0$-$202638}:  Cluster Abell S0026 is $50\arcsec$ from ID and overlaps within positional errors.
\\ \noindent {\bf 1RXS J002041.8$-$254307}:  Member of cluster Abell 0022.
\\ \noindent {\bf 1RXS J002136.8$+$280305}:  Group IV Zw 015 is $77\arcsec$ from ID and overlaps within positional errors.
\\ \noindent {\bf 1RXS J002430.5$-$292856}:  Extended radio emission.
\\ \noindent {\bf 1RXS J002534.9$-$330255}:  Greater than $3\sigma$ from X$-$ray position. Low X$-$ray
flux and large error region imply unreliable X$-$ray position. Optical and radio information point towards a correlation.
Member of cluster ABELL S0041.
\\ \noindent {\bf 1RXS J002811.6$+$310342}:  Possible ID is not at 002812.61$+$310315.3 (as identified in RGB). Better radio resolution would strengthen ID.
\\ \noindent {\bf 1RXS J003037.6$-$241053}:  Member of cluster Abell 0047.
\\ \noindent {\bf 1RXS J003413.7$-$212619}:  Member of group SCG 05.
\\ \noindent {\bf 1RXS J004324.9$-$203728}:  Member of cluster Abell 2813.
\\ \noindent {\bf 1RXS J004924.0$-$293128}:  Member of cluster Abell S0084.
\\ \noindent {\bf 1RXS J010721.2$+$322254}:  Greater than $3\sigma$ from X$-$ray position. Low X$-$ray
flux and large error region imply unreliable X$-$ray position. Optical and radio information point towards a correlation. Member of group IV Zw 038.
\\ \noindent {\bf 1RXS J010750.4$-$364342}:  Member of cluster Abell 2871.
\\ \noindent {\bf 1RXS J010850.5$-$152438}:  Member of cluster Abell 0151.
\\ \noindent {\bf 1RXS J011301.1$+$153041}:  Member of cluster Abell 0160. Given the low X$-$ray
flux and large error region imply unreliable X$-$ray position. Optical and radio information point towards a correlation.
\\ \noindent {\bf 1RXS J011354.9$-$314538}:  Greater than $3\sigma$ from X$-$ray position. Given the low X$-$ray
flux and large error region imply unreliable X$-$ray position. Optical and radio information point towards a correlation. 
Pointed {\it ASCA} observation by Arimoto et al. 1997 confirms ID. Member of cluster Abell S0141.
\\ \noindent {\bf 1RXS J011513.2$+$405743}:  More reliable ID is a blend of star and galaxy on DSS.
\\ \noindent {\bf 1RXS J012058.3$-$135100}:  Member of cluster CID 10. 
\\ \noindent {\bf 1RXS J012337.2$+$331506}:  Member of pair ARP 229.
\\ \noindent {\bf 1RXS J012507.3$+$084124}:  Member of cluster Abell 0193.
\\ \noindent {\bf 1RXS J012910.8$-$214158}:  Galaxy composed of two compact optical cores.
\\ \noindent {\bf 1RXS J013152.8$-$133651}:  Member of cluster Abell 0209.
\\ \noindent {\bf 1RXS J013632.9$+$390556}:  KPNO spectrum obtained. No features revealed amid strong blue continuum.
\\ \noindent {\bf 1RXS J013715.1$-$091145}:  Member of cluster CID 12.
\\ \noindent {\bf 1RXS J015244.7$+$360855}:  Member of cluster Abell 0262.
\\ \noindent {\bf 1RXS J015705.3$+$412029}:  Member of cluster Abell 0276.
\\ \noindent {\bf 1RXS J020144.2$-$021158}:  Member of cluster Abell 0291.
\\ \noindent {\bf 1RXS J023727.6$-$263027}:  Member of cluster Abell 0368.
\\ \noindent {\bf 1RXS J024122.1$-$283919}:  Cluster Abell 3041 is $53\arcsec$ from ID and overlaps within positional errors.
\\ \noindent {\bf 1RXS J024620.0$-$301639}:  Supernova SN 1992bd is $1\arcsec$ from ID and is also a valid identification. 
\\ \noindent {\bf 1RXS J024937.1$-$311114}:  Member of cluster Abell S0301.
\\ \noindent {\bf 1RXS J030150.6$+$355003}:  Member of group UGC 02489 and cluster Abell 0407.
V Zw 311 NOTES02 appears to be a cD from DSS and is likely to be true ID. Better radio resolution needed to clarify.
\\ \noindent {\bf 1RXS J031611.4$+$090445}:  KPNO spectrum obtained. No features revealed amid the flat continuum.
\\ \noindent {\bf 1RXS J033828.8$-$352701}:  Member of Fornax cluster.
\\ \noindent {\bf 1RXS J041325.4$+$102756}:  Member of cluster Abell 0478.
\\ \noindent {\bf 1RXS J042551.3$-$083329}:  Member of cluster EXO 0422$-$086.
\\ \noindent {\bf 1RXS J042700.9$-$204321}:  Member of cluster Abell 0490.
\\ \noindent {\bf 1RXS J043337.7$-$131528}:  Member of cluster Abell 0496.
\\ \noindent {\bf 1RXS J043643.8$+$100323}:  Member of cluster MS 0433.9$+$0957.
\\ \noindent {\bf 1RXS J043902.0$+$052050}:  Member of cluster RX J04390$+$0520.
\\ \noindent {\bf 1RXS J044511.8$-$155118}:  Member of cluster CID 25.
\\ \noindent {\bf 1RXS J044803.3$-$202741}:  High resolution radio position from Owen, White \& Ge (1993) confirms ID. Member of cluster Abell 0514.
\\ \noindent {\bf 1RXS J044814.8$+$095259}:  Member of cluster RX J0448.2+0952.
\\ \noindent {\bf 1RXS J045859.6$-$002904}:  Member of cluster CID 27.
\\ \noindent {\bf 1RXS J050049.4$-$384038}:  Member of cluster Abell 3301.
\\ \noindent {\bf 1RXS J050120.1$-$033223}:  High resolution radio position from Owen, White \& Ge (1993) confirms ID. Member of cluster Abell 0531. 
\\ \noindent {\bf 1RXS J050507.9$-$322007}:  Member of cluster Abell 3313.
\\ \noindent {\bf 1RXS J051611.5$-$000807}:  RBSC position incorrect. X-ray observations from \citet{ceballos96} confirm ID.
\\ \noindent {\bf 1RXS J054738.7$-$315228}:  Member of cluster Abell 3364.
\\ \noindent {\bf 1RXS J055040.7$-$321619}:  Cluster Abell S0549 ($z=0.04030$) is
$13\arcsec$ from ID and overlaps well within positional errors. 
\\ \noindent {\bf 1RXS J055711.1$-$372813}:  Member of cluster Abell S0555.
\\ \noindent {\bf 1RXS J060552.7$-$351759}:  Member of cluster Abell 3378.
\\ \noindent {\bf 1RXS J062707.5$-$352917}:  Member of cluster Abell 3392.
\\ \noindent {\bf 1RXS J064326.5$+$421420}:  High resolution radio position from \citet{sally97} confirms ID.
\\ \noindent {\bf 1RXS J070427.0$+$631856}:  Member of cluster Abell 0566.
\\ \noindent {\bf 1RXS J070907.6$+$483657}:  Member of cluster Abell 0569.
\\ \noindent {\bf 1RXS J071006.0$+$500243}:  High resolution radio position from \citet{sally97}.
\\ \noindent {\bf 1RXS J073221.5$+$313750}:  Member of cluster Abell 0586.
\\ \noindent {\bf 1RXS J074144.8$+$741444}:  Member of cluster ZwCl 0735.7$+$7421.
\\ \noindent {\bf 1RXS J074238.0$+$092213}:  Member of cluster Abell 0592. Better radio resolution needed.
\\ \noindent {\bf 1RXS J081021.3$+$421657}:  Member of cluster [VMF98] 047.
\\ \noindent {\bf 1RXS J081112.3$+$700300}:  Member of cluster Abell 0621
\\ \noindent {\bf 1RXS J081929.5$+$704221}:  Very extended radio and
optical emission associated with the dwarf galaxy Holmberg II.
\\ \noindent {\bf 1RXS J082209.5$+$470601}:  Member of cluster Abell 0646.
\\ \noindent {\bf 1RXS J083811.0$+$245336}:  Member of pair CGCG 120$-$011.
\\ \noindent {\bf 1RXS J083950.7$-$121424}:  High resolution radio position from \citet{li96} confirm ID as [HB89] 0837-120 and not [EYG89] 087.
\\ \noindent {\bf 1RXS J084255.9$+$292752}:  Member of cluster ZwCl 0839.9$+$2937.
\\ \noindent {\bf 1RXS J091037.2$+$332920}:  KPNO spectrum obtained. No features revealed among strong blue continuum.
\\ \noindent {\bf 1RXS J091949.1$+$334532}:  X-ray position offset, but Einstein observations from Fabbiano, Kim \& Trinchieri (1992) confirm ID. 
Member of cluster Abell 0779.
\\ \noindent {\bf 1RXS J092406.1$+$141002}:  Member of cluster Abell 0795.
\\ \noindent {\bf 1RXS J094740.4$-$305655}:  Member of triple galaxy AM 0945$-$304.
\\ \noindent {\bf 1RXS J095821.8$-$110344}:  Member of cluster Abell 0907.
\\ \noindent {\bf 1RXS J100121.5$+$555351}:  Member of cluster [YGK80] 0957$+$561CLUSTER. QSO [HB89] 0957$+$561
and galaxy [YGK81] 097 are $4\arcsec$ and $9\arcsec$ from ID and overlap within positional errors.
\\ \noindent {\bf 1RXS J100639.1$+$255450}:  Member of cluster Abell 0923.
\\ \noindent {\bf 1RXS J101335.9$-$135108}:  Member of cluster RX J10136$-$1350.
\\ \noindent {\bf 1RXS J101912.1$+$635802}:  Member of pair CGCG 313$-$011.
\\ \noindent {\bf 1RXS J102228.9$+$500630}:  Member of cluster Abell 0980.
\\ \noindent {\bf 1RXS J102339.3$+$041117}:  Member of cluster ZwCl 1021.0$+$0426.
\\ \noindent {\bf 1RXS J102350.7$-$271522}:  Member of cluster Abell 3444.
\\ \noindent {\bf 1RXS J102758.9$-$064804}:  Member of cluster RX J1027.9-0648.
\\ \noindent {\bf 1RXS J103155.8$-$141659}:  Multiple redshift systems overlap 
within positional errors.
\\ \noindent {\bf 1RXS J103459.5$+$304138}:  Member of cluster Abell 1045.
\\ \noindent {\bf 1RXS J104043.7$+$395706}:  Member of cluster Abell 1068.
\\ \noindent {\bf 1RXS J104431.7$-$070404}:  Member of cluster Abell 1084.
\\ \noindent {\bf 1RXS J104651.9$-$253546}:  Member of cluster RX J10468$-$2535
\\ \noindent {\bf 1RXS J105344.2$+$492956}:  Member of cluster MS 1050.7$+$4946.
\\ \noindent {\bf 1RXS J105825.9$+$564716}:  Member of cluster Abell 1132.
\\ \noindent {\bf 1RXS J111137.2$+$405031}:  Member of cluster Abell 1190.
\\ \noindent {\bf 1RXS J111422.6$+$582318}:  High resolution radio position from
\citet{sally97} confirms ID.
\\ \noindent {\bf 1RXS J111450.1$-$121351}:  Cluster Abell 1211 is $89\arcsec$ from ID and overlaps within positional errors.
\\ \noindent {\bf 1RXS J113121.4$+$333447}:  Member of cluster Abell 1290.
\\ \noindent {\bf 1RXS J113153.7$-$195543}:  Member of cluster Abell 1300.
\\ \noindent {\bf 1RXS J113222.4$+$555828}:  Galaxy MCG $+$09$-$19$-$110 ($z=0.05130$) is $17\arcsec$ from ID and
overlaps within positional errors. Member of cluster RX J1132.3$+$5558.
\\ \noindent {\bf 1RXS J113448.4$+$490438}:  Member of cluster Abell 1314.
\\ \noindent {\bf 1RXS J114124.2$-$121632}:  Member of cluster RX J1141.4$-$1216.
\\ \noindent {\bf 1RXS J114442.2$+$672435}:  Member of cluster Abell 1366. Optical and radio information point towards a correlation. Possible X-ray cluster emission?
\\ \noindent {\bf 1RXS J114452.7$+$194706}:  Greater than $3\sigma$ from RBSC position. Public {\it ROSAT} HRI observations
place X$-$ray source much closer to reliable ID. Association with cluster Abell 1367.
\\ \noindent {\bf 1RXS J114947.0$-$121845}:  Member of Abell 1391. Probable association.
\\ \noindent {\bf 1RXS J115518.9$+$232431}:  Member of cluster Abell 1413.
taken from the FIRST 1.4 GHz survey \citep{white97}. NVSS flux density is 5.4 mJy. Member of cluster RX J11570$+$2415.
\\ \noindent {\bf 1RXS J115719.0$+$333645}:  High resolution radio position from Owen, White \& Ge (1993) confirms more reliable ID. Member of cluster Abell 1423. 
\\ \noindent {\bf 1RXS J120511.7$+$392043}:  Member of cluster RX J1205.1$+$3920
\\ \noindent {\bf 1RXS J121105.4$+$352005}:  Member of cluster RX J1211.0$+$3520
\\ \noindent {\bf 1RXS J122752.7$+$632317}:  Member of cluster RX J1227.8$+$6323.
\\ \noindent {\bf 1RXS J122945.9$+$075927}:  Member of cluster Virgo S.
\\ \noindent {\bf 1RXS J123658.8$+$631111}:  Member of cluster Abell 1576.
\\ \noindent {\bf 1RXS J124349.3$-$153320}:  Member of cluster Abell 1603.
\\ \noindent {\bf 1RXS J125233.7$-$311605}:  Member of cluster RX J12525$-$3116.
\\ \noindent {\bf 1RXS J125422.2$-$290034}:  Member of cluster Abell 3528.
\\ \noindent {\bf 1RXS J125710.0$-$172345}:  Pointed {\it ROSAT} observation from \citet{peres98} confirms ID. Member of cluster Abell 1644. 
\\ \noindent {\bf 1RXS J125921.5$-$041131}:  Member of cluster Abell 1651.
\\ \noindent {\bf 1RXS J130250.3$-$023041}:  Member of cluster Abell 1663.
\\ \noindent {\bf 1RXS J130343.6$-$241506}:  Member of cluster Abell 1664.
\\ \noindent {\bf 1RXS J130346.7$+$191635}:  Member of cluster Abell 1668.
\\ \noindent {\bf 1RXS J130552.6$+$305405}:  Member of cluster Abell 1677.
\\ \noindent {\bf 1RXS J130916.1$-$013658}:  Member of cluster MS 1306.7$-$0121.
\\ \noindent {\bf 1RXS J131129.5$-$012017}:  Member of cluster Abell 1689.
\\ \noindent {\bf 1RXS J131506.8$+$514931}:  Member of cluster Abell 1703.
\\ \noindent {\bf 1RXS J132016.3$+$330828}:  Member of cluster RX J1320.1$+$3308.
\\ \noindent {\bf 1RXS J132542.1$-$022800}:  No optical counterpart on DSS.
\\ \noindent {\bf 1RXS J132549.2$+$591937}:  Member of cluster Abell 1744.
\\ \noindent {\bf 1RXS J132617.4$+$001329}:  Member of cluster RX J1326.3$+$0013.
\\ \noindent {\bf 1RXS J132758.9$-$313027}:  Greater than $3\sigma$ from RBSC position. 
Optical and radio information point towards a correlation. Member of cluster Abell
3558. Possible X-ray blend of star and galaxy?
\\ \noindent {\bf 1RXS J133226.0$-$330812}:  Better radio resolution confirms ID.
\\ \noindent {\bf 1RXS J134104.8$+$395942}:  Member of cluster Abell 1774.
\\ \noindent {\bf 1RXS J134152.6$+$262230}:  Cluster Abell 1775 ($z=0.06960$) is
$72\arcsec$ from ID and overlaps within positional errors.
\\ \noindent {\bf 1RXS J134730.5$-$114455}:  Member of cluster RX J13475$-$1145.
\\ \noindent {\bf 1RXS J134852.6$+$263541}:  Member of cluster Abell 1795.
\\ \noindent {\bf 1RXS J140102.1$+$025249}:  Member of cluster Abell 1835.
\\ \noindent {\bf 1RXS J140337.0$-$335840}:  Member of cluster Abell S0753.
\\ \noindent {\bf 1RXS J140728.4$-$270055}:  Member of cluster Abell 3581.
\\ \noindent {\bf 1RXS J141342.6$+$433938}:  Member of cluster Abell 1885.
\\ \noindent {\bf 1RXS J141357.3$+$711751}:  Association with cluster Abell 1895 ($z=0.22492$).
\\ \noindent {\bf 1RXS J142139.7$+$371743}:  Member of cluster Abell 1902.
\\ \noindent {\bf 1RXS J143236.0$+$313855}:  Cluster Abell 1930 ($z=0.13130$) is $119\arcsec$ from ID and overlaps within positional errors.
\\ \noindent {\bf 1RXS J143527.9$+$550747}:  Member of cluster Abell 1940.
\\ \noindent {\bf 1RXS J144428.4$+$311304}:  Cluster Abell 1961 ($z=0.23200$) is $117\arcsec$ from ID and overlaps within positional errors.
\\ \noindent {\bf 1RXS J145307.8$+$215333}:  Member of cluster Abell 1986.
\\ \noindent {\bf 1RXS J145431.4$+$183834}:  Member of cluster Abell 1991.
\\ \noindent {\bf 1RXS J145434.1$+$080250}:  Greater than $3\sigma$ from RBSC position. \citet{bade98} claim optical$-$$-$radio ID is 
the correct X$-$ray source. Optical and radio information point towards a correlation. 
\\ \noindent {\bf 1RXS J145507.9$+$192025}:  High resolution radio position from
\citet{sally97} confirms ID.
\\ \noindent {\bf 1RXS J145715.4$+$222026}:  Member of cluster MS 1455.0$+$2232.
\\ \noindent {\bf 1RXS J145904.1$-$084254}:  Better radio resolution needed to confirm ID.
\\ \noindent {\bf 1RXS J150020.7$+$212213}:  Member of cluster Abell 2009.
\\ \noindent {\bf 1RXS J151056.3$+$054431}:  Member of cluster Abell 2029.
\\ \noindent {\bf 1RXS J151127.2$+$062153}:  Greater than $3\sigma$ from RBSC 
position. Optical and radio information point towards a correlation. 
Member of cluster Abell 2033. Possible X-ray cluster emission?
\\ \noindent {\bf 1RXS J151642.4$+$070058}:  Member of cluster Abell 2052.
\\ \noindent {\bf 1RXS J151845.3$+$061340}:  High resolution radio position from 
\citet{sally97} confirms ID. Member of cluster Abell 2055.
\\ \noindent {\bf 1RXS J152151.0$+$074221}:  Member of cluster MKW 03s.
\\ \noindent {\bf 1RXS J152305.7$+$083550}:  ID greater than $3\sigma$ of RBSC 
position. There is a strong optical/radio candidate (CGCG 077-097) located 
within the cluster Abell 2063.
\\ \noindent {\bf 1RXS J153950.3$+$304305}:  Member of cluster Abell 2110.
\\ \noindent {\bf 1RXS J154009.4$+$141116}:  High resolution radio position 
from \citet{sally97} confirms ID (a star?).
\\ \noindent {\bf 1RXS J155611.0$+$662123}:  Association with cluster Abell 2146.
\\ \noindent {\bf 1RXS J160435.1$+$174333}:  High resolution X-ray position from 
\citet{burstein97} confirms ID.
\\ \noindent {\bf 1RXS J160456.8$+$235604}:  Member of cluster AWM 4.
\\ \noindent {\bf 1RXS J160740.7$+$254106}:  Optical ID is offset from strong radio source (FIRST ID), possibly
indicative of a misidentification with a foreground star.
\\ \noindent {\bf 1RXS J161546.9$-$060841}:  Radio emission is very diffuse, from multiple components. Better radio
resolution needed. Association.
\\ \noindent {\bf 1RXS J161711.4$+$063816}:  No object found within $3\sigma$ of X$-$ray source $-$ X$-$ray errors too
conservative? Identified as RGB J1617$+$066. Optical and radio information point towards a correlation. 
High resolution radio position from \citet{sally97}. 
\\ \noindent {\bf 1RXS J162032.0$+$295321}:  Member of cluster Abell 2175.
\\ \noindent {\bf 1RXS J162100.4$+$254547}:  Member of Abell 2177.
\\ \noindent {\bf 1RXS J162837.7$+$393249}:  Member of cluster Abell 2199.
\\ \noindent {\bf 1RXS J163246.8$+$053423}:  Member of cluster Abell 2204
\\ \noindent {\bf 1RXS J164022.1$+$464215}:  Member of cluster Abell 2219
\\ \noindent {\bf 1RXS J164238.9$+$272621}:  Member of cluster Abell 2223.
\\ \noindent {\bf 1RXS J170242.5$+$340336}:  Member of cluster Abell 2244
\\ \noindent {\bf 1RXS J170941.2$+$342529}:  Member of cluster Abell 2249.
\\ \noindent {\bf 1RXS J171519.5$+$572430}:  Member of cluster CID 71.
\\ \noindent {\bf 1RXS J171706.8$+$293117}:  Member of cluster RBS 1634.
\\ \noindent {\bf 1RXS J171718.9$+$422652}:  Member of cluster RX J17173$+$4227.
\\ \noindent {\bf 1RXS J171746.9$+$194057}:  Association with cluster Abell 2254.
\\ \noindent {\bf 1RXS J171810.9$+$563955}:  Member of cluster ZwCl 1717.9$+$5636.
\\ \noindent {\bf 1RXS J172009.3$+$263727}:  Member of cluster RX J17201$+$2637.
\\ \noindent {\bf 1RXS J172226.7$+$320752}:  Member of cluster Abell 2261.
\\ \noindent {\bf 1RXS J173257.0$+$403635}:  Association with cluster Abell 2272.
\\ \noindent {\bf 1RXS J173301.7$+$434533}:  Member of cluster CID 72.
\\ \noindent {\bf 1RXS J174414.3$+$325925}:  Member of cluster ZwCl 1742.1$+$3306.
\\ \noindent {\bf 1RXS J175004.4$+$470037}:  High resolution radio position
from \citet{sally97} confirms ID.
\\ \noindent {\bf 1RXS J175441.9$+$680334}:  Cluster ZwCl 1754.5$+$6807 ($z=0.07700$) is $53\arcsec$ from ID and
overlaps within positional errors.
\\ \noindent {\bf 1RXS J175706.9$+$535130}:  Member of cluster Abell 2292.
\\ \noindent {\bf 1RXS J182157.4$+$642051}:  High resolution radio position from 
\citet{sally97} confirms ID.
\\ \noindent {\bf 1RXS J203445.3$-$354921}:  Member of cluster Abell 3695.
\\ \noindent {\bf 1RXS J210027.7$-$170913}:  Better radio resolution needed to confirm ID.
\\ \noindent {\bf 1RXS J210707.7$-$252643}:  Cluster Abell 3744 is nearby and provides a considerable amount of diffuse radio emission as well. Better radio resolution needed.
\\ \noindent {\bf 1RXS J212706.7$-$120927}:  Greater than $3\sigma$ of RBSC 
position. Optical and radio information point towards a correlation. 
\\ \noindent {\bf 1RXS J213414.4$-$311737}:  Better radio resolution needed to confirm ID.
\\ \noindent {\bf 1RXS J213504.7$+$085807}:  Optical ID is actually a composite of a bright star and another 
fainter source.
\\ \noindent {\bf 1RXS J214015.3$-$233946}:  Member of cluster MS 2137.3$-$2353.
\\ \noindent {\bf 1RXS J215336.1$+$174111}:  Member of cluster Abell 2390.
\\ \noindent {\bf 1RXS J215541.5$+$123144}:  Member of cluster Abell 2396
\\ \noindent {\bf 1RXS J221020.8$-$121040}:  Member of cluster Abell 2420.
\\ \noindent {\bf 1RXS J221614.9$-$092033}:  Greater than $3\sigma$ of RBSC 
position. Optical and radio information point towards a correlation.
Possible X-ray cluster emission?
\\ \noindent {\bf 1RXS J224919.5$-$372724}:  Member of cluster Abell S1065.
\\ \noindent {\bf 1RXS J225019.2$+$105406}:  Member of cluster Abell 2495.
\\ \noindent {\bf 1RXS J225334.1$-$334309}:  Member of cluster Abell 3934.
\\ \noindent {\bf 1RXS J230714.9$-$151322}:  Member of cluster Abell 2533.
\\ \noindent {\bf 1RXS J231607.5$-$202739}:  Member of cluster Abell 2566.
\\ \noindent {\bf 1RXS J231712.8$+$184237}:  Member of group HCG 094.
\\ \noindent {\bf 1RXS J231829.5$+$184246}:  Greater than $3\sigma$ of RBSC 
position. Optical and radio information point towards a correlation.
Member of cluster Abell 2572. Possible X-ray cluster emission?
\\ \noindent {\bf 1RXS J232125.9$-$231230}:  Member of cluster Abell 2580.
\\ \noindent {\bf 1RXS J232519.4$-$120741}:  Member of cluster Abell 2597.
\\ \noindent {\bf 1RXS J233631.0$+$210848}:  Member of cluster Abell 2626.
\\ \noindent {\bf 1RXS J233641.8$+$235526}:  Member of cluster Abell 2627.
\\ \noindent {\bf 1RXS J233827.6$+$270045}:  Greater than $3\sigma$ of RBSC 
position. Optical and radio information point towards a correlation.
Member of cluster Abell 2634. Pointed {\it ROSAT} HRI observations by
Sakelliou \& Merrifield 1998 show X-ray emission is a blend of cluster
and individual galaxy emission. 
\\ \noindent {\bf J234741.2$-$280829}:  X-ray emission appears to be
from bright knot in radio jet. Optical and radio information point towards a correlation.
\\ \noindent {\bf 1RXS J235034.7$+$292924}:  Cluster MS 2348.0$+$2913 ($z=0.09500$) is $22\arcsec$ from ID and
overlaps within positional errors. KPNO spectrum obtained. Strong absorption features at $z=$-$0.00020$, possibly a foreground star amid the
cluster. Better optical and radio resolution needed to clarify.
\\ \noindent {\bf 1RXS J234741.2$-$280829}:  ID greater than $3\sigma$ of RBSC 
position. There is a strong optical/radio candidate nearby. Member of cluster Abell 4038.

\normalsize

\subsection{Previous RBSC Misidentifications?}

\scriptsize 

\noindent {\bf 1RXS J000856.1$+$411034}:  Better radio resolution
needed. Possible X-ray ID does not seem to be B3 0006$+$408 (as
identified in RGB), but rather a bright, diffuse object at 00h08m56.0s $+$41d10m09.1s. 
\\ \noindent {\bf 1RXS J001144.5$+$322441}:  Radio ID is 87GB 000913.4$+$320913 (as identified in RGB), and X$-$ray ID is cluster Abell 0007.
\\ \noindent {\bf 1RXS J220220.8$+$035306}: Optical/X$-$ray ID greater
than $3\sigma$ from radio position (NVSS and RGB). Background radio
object misidentified with B=13.2 X-ray star? 

\normalsize

\subsection{A simple, well-defined sample of AGN}\label{discussion}

The RBSC-NVSS sample is defined by the following criteria:
\renewcommand{\baselinestretch}{1.2}
\begin{itemize}
\item RBSC count rate $\ge$ 0.1 counts s$^{-1}$
\item $f_{r} \ge 2.5 mJy$
\item $\delta \ge -40^\circ$
\item $|b| \ge 15^\circ$
% \item B $\le 22$
\end{itemize}
\normalsize

\noindent Columns 1 and 2 of Table~\ref{process} summarize the
results of the selection process. We should note here, that while the
method we employ here is quite robust, the large number of RBSC
sources and the loose positional errors dictate that there will be
some spurious spatial coincidences. To estimate the number of false
matches we might have, we shifted the X-ray positions by $8\arcmin$ and
performed the identification procedure again. The results are listed
in column 3 of Table~\ref{process}, where we find that the expected
percentage of spurious matches is around 3$\%$ for reliabilities above 
0.50. This fraction could be further reduced by comparing the flux
density and magnitude distributions of the true and spurious objects;
most spurious sources lie near the NVSS and USNO survey limits, while
RBSC-NVSS sources do not.

Forthcoming papers in this series will present the
radio, optical, and X-ray properties of the sample and optical
identifications and classifications of a subset of objects. Using a
number of criteria, we find that this sample of 1512 extragalactic
objects is comprised almost entirely of AGN, making this the largest,
complete sample of its kind. It represents a major step forward in the 
identification of RBSC objects and contains a large sample of both
radio-loud {\it and} radio-quiet X-ray objects (previous surveys of
this type have typically sampled only the radio-loud population).

\placetable{process}

\acknowledgements

FEB thanks W. Brinkmann and J. Siebert for their helpful comments and
assistance with the RBSC. He also acknowledges support from a National
Radio Astronomy Observatory Jansky Pre-doctoral Fellowship. JJB
acknowledges support from the NSF through grant AST 9320547.
This research has made use of data obtained through 
the High Energy Astrophysics Science Archive Research Center Online Service, 
provided by the NASA/Goddard Space Flight Center, the
NASA/IPAC Extragalactic Database (NED) which is operated by the Jet
Propulsion Laboratory, Caltech, under contract with the National
Aeronautics and Space Administration and the SIMBAD database,
operated at CDS, Strasbourg, France. 
The Digitized Sky Surveys were produced at the Space Telescope Science
Institute under US government grant NAG W-2166. 

\clearpage
\begin{deluxetable}{llclrrlrl}
\tabletypesize{\scriptsize}
\tablewidth{0pt}
\tablecaption{RBSC Galactic Identifications \label{stars}} 
\tablehead{
\colhead{(1)} & \colhead{(2)} & \colhead{(3)} & 
\colhead{(4)} & \colhead{(5)} & \colhead{(6)} & 
\colhead{(7)} & \colhead{(8)} & \colhead{(9)} \ 
\\ 
\colhead{{\it ROSAT} BSC} & \colhead{SIMBAD} & \colhead{S$_{\rm x}$} & 
\colhead{Position} & \colhead{S$_{\rm 1.4 GHz}$} & 
\colhead{ID} & \multicolumn{3}{c}{SIMBAD} \ 
\\ 
\cline{7-9} 
\colhead{1RXS name} & \colhead{Common name} & \colhead{{\rm {\llap er}gs s$^{-1}
$ c{\rlap m\ \ $^{-2}$}}} & 
\colhead{$\alpha$ \quad (J2000) \quad $\delta$} & \colhead{{\rm mJy}} & 
\colhead{P} & \colhead{Type} & \colhead{m$_{B}$} &  
\colhead{Sp. Class.} \ 
}
\tableheadfrac{0.03}
\startdata
J001208.2$-$155031 & SAO 147143                    & 1.52e$-$11 & 00 12 07.78 $-$15 50 38.6     &     2.6  & 0.94 & *     &  10.5 & K0       \\
J004703.1$-$115217 & NGC 0246                      & 1.01e$-$11 & 00 47 03.41 $-$11 52 19.1\tablenotemark{O}    &   126.7\tablenotemark{E} & 1.00 & PN    &  15.8\tablenotemark{~} &         \\
J020135.5$-$160957 & HD 12538                      & 1.60e$-$12 & 02 01 35.23 $-$16 10 19.4     &     3.2  & 1.00 & *     &   7.1 & A3      \\
J031634.9$+$321115 & HD 20277                      & 2.75e$-$11 & 03 16 34.00 $+$32 11 15.9     &     2.8  & 0.98 & *sr   &   7.0 & G8IV     \\
J032635.1$+$284302 & UX Ari                        & 2.65e$-$10 & 03 26 35.40 $+$28 42 54.2     &    41.6  & 1.00 & *RS   &   6.5 &          \\
J033647.2$+$003518 & RGB J0336$+$005               & 7.10e$-$10 & 03 36 47.25 $+$00 35 15.5     &   123.2  & 1.00 & *RS   &   5.7 & G9    \\
J040940.8$-$075327 & V* EI Eri                     & 1.34e$-$10 & 04 09 40.92 $-$07 53 31.7     &     3.9  & 1.00 & *RS   &   7.7 & G5IV     \\
J041411.9$+$281230 &  HD 283447                    & 5.86e$-$12 & 04 14 13.12 $+$28 12 14.8     &     6.0  & 1.00 & *TT   &  11.8 & K2        \\
J052507.7$+$062103 & SV* ZI 374                    & 9.87e$-$12 & 05 25 08.18 $+$06 20 50.6     &     3.0  & 0.99 & *V    &   1.4 & B2III     \\
J053516.6$-$052320 & M42                           & 1.30e$-$10 & 05 35 17.17 $-$05 22 32.7     & 105702.0  & 0.97 & *Cl   &   3.0 & HII      \\
J062150.0$-$341140 & V* HY CMa                     & 3.57e$-$12 & 06 21 49.55 $-$34 11 54.1     &     8.6  & 1.00 & *V    &  10.3 & K0        \\
J072132.9$+$260939 & V* V340 Gem                   & 1.01e$-$11 & 07 21 32.89 $+$26 09 27.6     &     7.5  & 0.95 & *sr   &   8.4 & G2V        \\
J072931.4$+$355607 &                               & 6.47e$-$12 & 07 29 31.76 $+$35 55 51.6     &     7.3  & 0.99 & *     &  (12.9{\rlap)}&          \\
J074318.7$+$285306 & HD 62044                      & 2.15e$-$10 & 07 43 18.53 $+$28 53 02.5     &     6.5  & 0.98 & *RS  &   4.3 & K1III    \\
J093346.5$+$624943 & V* FF UMa                     & 4.49e$-$11 & 09 33 46.94 $+$62 49 40.0     &     7.4  & 1.00 & *V    &   8.9 & G5        \\
J103105.7$+$823327 & NSV 4864                      & 7.81e$-$12 & 10 31 05.02 $+$82 33 30.7\tablenotemark{O}    &     9.6  & 0.93 & *V    &   5.6 & F2V        \\
J104520.5$+$453404 & V* TX UMa                     & 2.26e$-$12 & 10 45 20.50 $+$45 33 58.7\tablenotemark{O}    &     9.6  & 0.71 & *Al   &   7.1 & B8V       \\
J113155.7$-$343632 & CD$-$33 7795                  & 1.93e$-$11 & 11 31 55.14 $-$34 36 29.4     &     5.6  & 0.97 & *     &  (11.8{\rlap)}& M1       \\
J115800.5$+$140222 & V* FZ Leo                     & 1.45e$-$11 & 11 58 02.16 $+$14 02 18.5     &     2.9  & 0.95 & *V    &   8.9 & G5          \\
J133447.5$+$371100 & V* BH CVn                     & 2.78e$-$11 & 13 34 47.62 $+$37 10 53.8     &     8.8  & 0.91 & *RS   &   5.3 & F2IV         \\
J142555.6$+$141148 & StKM 1$-$1155                 & 5.16e$-$12 & 14 25 56.00 $+$14 12 14.0\tablenotemark{O}    &     2.8  & 0.67 & *     &  11.4 & K4          \\
                   &                               &            & 14 25 56.16 $+$14 12 02.1\tablenotemark{O}    &     2.8  & 0.26 &       &  (19.1{\rlap)}&          \\
J143638.2$+$584309 &                               & 1.41e$-$12 & 14 36 36.86 $+$58 42 55.8     &     4.5  & 1.00 & *     &  (11.1{\rlap)}&          \\
J150058.0$-$083103 & V* del Lib                    & 2.22e$-$11 & 15 00 58.38 $-$08 31 10.2     &     4.8  & 0.98 & *Al   &   4.9 & B9.5V      \\
J151720.9$-$190103 & HD 135743                     &  1.39e$-$11 & 15 17 21.22 $-$19 00 58.9\tablenotemark{O}    &     3.3  & 0.93 & *     &  10.0 & G3/G5V        \\
                   &                               &         & 15 17 21.26 $-$19 01 12.1\tablenotemark{O}    &     3.3  & 0.07 &       &  (17.2{\rlap)}&          \\
J152153.0$+$205830 & V* OT Ser                     & 2.65e$-$11 & 15 21 52.98 $+$20 58 38.1     &    15.4  & 0.98 & *V    &  11.4 & M9           \\
J154131.4$-$252043 & CD$-$24 12231                 & 1.39e$-$11 & 15 41 31.23 $-$25 20 36.2\tablenotemark{O}    &     3.0  & 0.83 & *     &  10.7 & K0        \\
J161216.4$-$282500 & SV* ZI 1217                   & 8.58e$-$12 & 16 12 16.34 $-$28 25 00.9\tablenotemark{O}    &     5.9  & 0.76 & *V    &   5.7 & B9Vvar       \\
J161441.0$+$335125 & HD 146361 J                   & 1.32e$-$10 & 16 14 41.01 $+$33 51 43.5     &     4.6  & 1.00 & *RS  &   5.6 & G0Ve     \\
J165529.3$-$082008 & V* V1054 Oph                  & 1.85e$-$10 & 16 55 28.89 $-$08 20 10.2     &    10.7  & 1.00 & *Fl   &  10.6 & M3Ve        \\
J172839.3$+$590158 & V* GR Dra                     & 1.92e$-$11 & 17 28 39.36 $+$59 02 07.0\tablenotemark{O}    &     3.3  & 0.87 & *Pu   &   8.8 & G0           \\
J173241.3$+$741337 & V* DR Dra                     & 3.13e$-$11 & 17 32 40.45 $+$74 13 32.8     &     3.1  & 0.99 & *RS   &   7.7 & K0III       \\
J173734.1$+$414618 &                               & 2.11e$-$12 & 17 37 32.90 $+$41 46 12.4     &    10.4  & 0.99 & *     &  (12.5{\rlap)}&          \\
J204009.4$-$005216 & V* AE Aqr                     & 2.94e$-$11 & 20 40 09.02 $-$00 52 15.5\tablenotemark{O}    &     4.3  & 0.93 & *DQ   &  10.4 & K5IV$-$Vvar     \\
                   &                               &              & 20 40 08.92 $-$00 52 23.7\tablenotemark{O}    &     4.3  & 0.05 &       &  (18.7{\rlap)}&          \\
J204151.2$-$322604 & V* AT Mic                     & 1.15e$-$10 & 20 41 51.08 $-$32 26 03.2     &     3.3  & 0.99 & *V    &  11.8 & M4.5         \\
J204512.3$-$350953 & V* BX Mic                     & 1.93e$-$11 & 20 45 11.79 $-$35 10 01.0     &     3.8  & 1.00 & *bL   &   7.4 &  G0V         \\
J210952.2$+$160920 & V* NR Peg                     & 4.20e$-$11 & 21 09 52.10 $+$16 09 26.7    &     2.7  & 0.78 & *bL   &    8.7 &  G0        \\
J212958.4$+$120959 & NGC 7078                      & 1.53e$-$10 & 21 29 59.27 $+$12 10 14.1     &     6.3  & 0.98 & *Cl   &  $-\ $ &          \\
J221719.9$-$084801 & GJ 852                        & 6.98e$-$12 & 22 17 20.32 $-$08 48 02.2\tablenotemark{O}    &     7.0  &  $-\ $ & *     &  (15.9{\rlap)}& M4.5Vmevar  \\
J225302.0$+$165029 & V* IM Peg                     & 8.33e$-$11 & 22 53 02.39 $+$16 50 29.4\tablenotemark{O}    &     3.0  & 0.87 & *RS   &   7.0 & K1.5II$-$IIIe     \\
J232301.2$-$063545 & HD 220338                     & 1.74e$-$11 & 23 23 01.09 $-$06 35 45.8     &    13.4  & 1.00 & *     &  10.2 & K0          \\
J233152.6$+$195735 & V* EQ Peg                     & 2.62e$-$11 & 23 31 51.81 $+$19 56 15.2     &     7.8  & $-\ $  & *     &  11.5 & M3.5        \\ 
J234351.0$-$151655 & V* R Aqr                      & 1.73e$-$12 & 23 43 49.56 $-$15 17 03.7     &    18.4  & 0.99 & *Mi   &   8.8 & M7IIIpevar      \\
J234940.8$+$362525 & AG$+$36 2436                  & 2.38e$-$11 & 23 49 41.00 $+$36 25 30.8     &    18.1  & 1.00 & *Ro   &   6.7 & G1IIIe       \\
\enddata
\tablenotetext{O}{Optical position given: because identification is more than 3$
\sigma$ from radio position or there are multiple optical candidates}
\tablenotetext{E}{Extended radio emission.}
\end{deluxetable}

\clearpage
\begin{table}
\dummytable\label{gals}
\end{table}

\clearpage
\begin{deluxetable}{lcc}
\tabletypesize{\small}
\tablewidth{480pt}
\tablecaption{RBSC-NVSS Results \label{process}} 
\tablehead{
\colhead{Selection} & \colhead{No. of} & \colhead{No. of spurious} \
\\ 
\colhead{criteria} & \colhead{identifications} & \colhead{identifications} \
\\ 
\colhead{imposed} & \colhead{found} & \colhead{expected} \
\\ 
}
\startdata
RBSC sample & 18811 & $-$ \\
Count rate $\ge$ 0.1 count s$^{-1}$ & 8547 & $-$ \\
$|b| \ge 15^{\circ}$, $\delta \ge -40^{\circ}$ & 5441 & $-$\\
RBSC-NVSS matches with $\Delta$ position $< 3\sigma$ and $f_{r} \ge
2.5$ mJy & 2300 & 919 \\
Linked cross-identification matches & 1556 & 51\tablenotemark{*} \\
\enddata
\tablenotetext{*}{The flux density and magnitude distributions of the
spurious identifications are markedly different from the RBSC-NVSS
sample, } 
\end{deluxetable}

\clearpage

%
%	Figure Captions
%

\figcaption[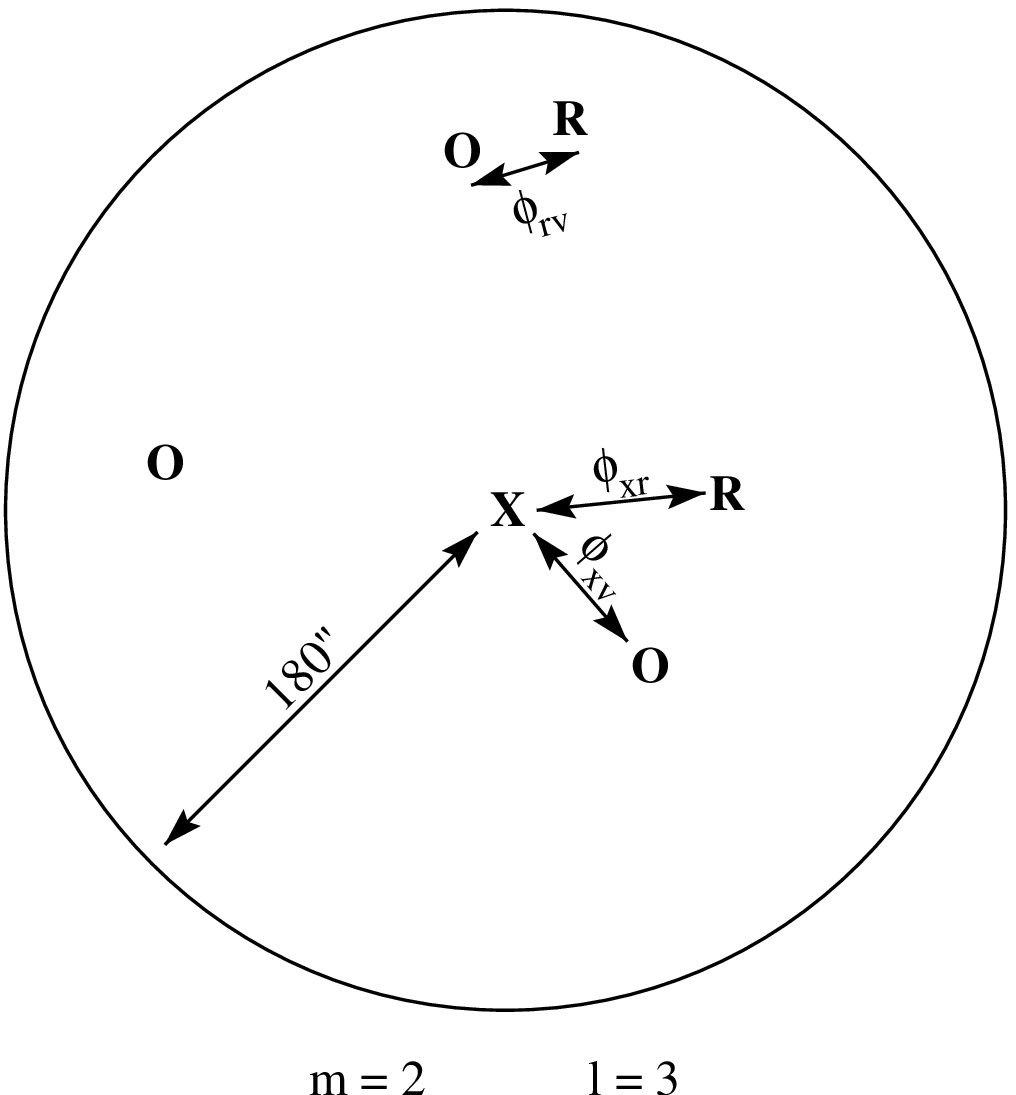]{
Diagram of a typical RBSC-NVSS field. Each search area contains
some numbers $m \geq 0$ of NVSS radio sources and $l \geq 0$ of
optical objects from the USNO catalog. In this case, there are three
optical sources (O) and two radio sources (R) positionally offset from 
the X-ray source (X) by varying angles ($\phi$).
\label{fig:diagram}}

\figcaption[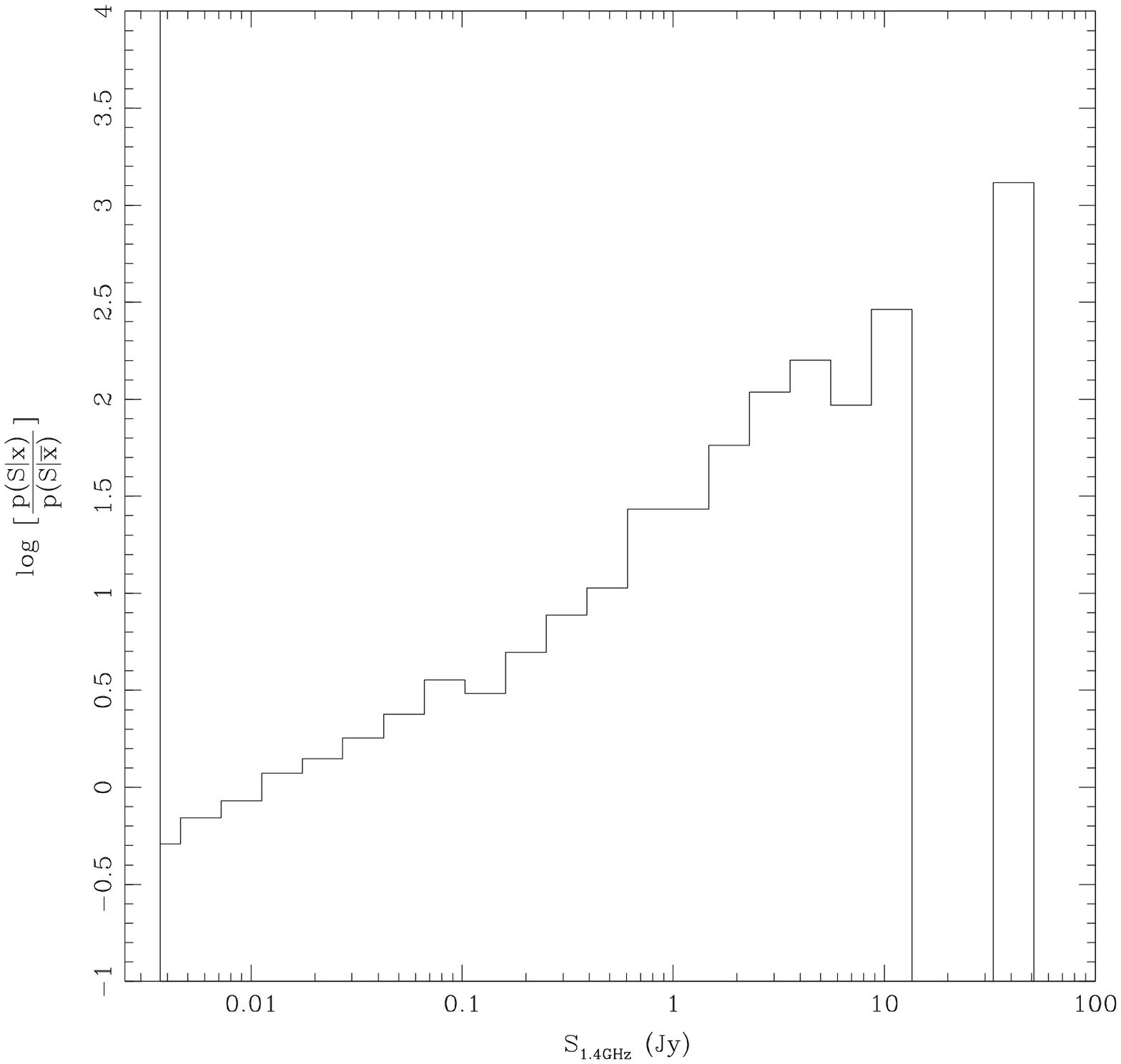]{
Logarithm of the ratio of the probability $p[S\vert x]$ that a
RBSC-NVSS source has flux density $S$ to the probability $p[S\vert
\overline{x}]$ that an unrelated NVSS sources has flux density $S$.
This plot indicates that most background radio sources are much
fainter than the true radio identifications of X-ray sources and can
be exploited to enhance the probabilities of identifications of strong
radio candidates.
\label{fig:radsrc}}
    
\figcaption[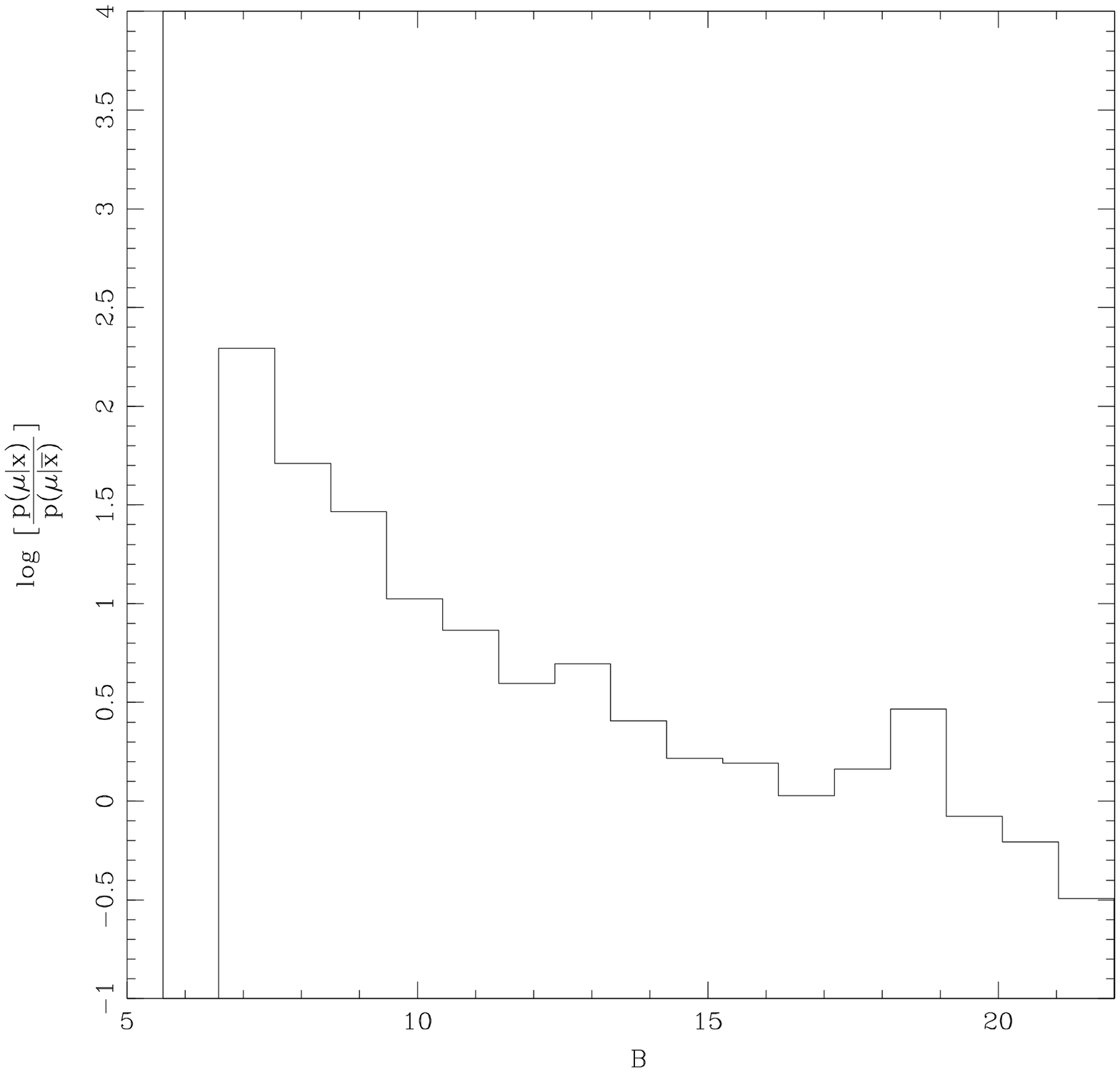]{
Logarithm of the ratio of the probability $p[\mu\vert x]$ that a
RBSC-NVSS optical identification has magnitude $\mu$ to the
probability $p[\mu\vert \overline{x}]$ that an unrelated USNO object
\citep{bahcall80} has magnitude $\mu$. This plot demonstrates that the
magnitude distribution of background optical sources is similar to the
optical identifications of RBSC sources, yielding little or no
enhancement of the probabilities of optical counterparts except at
bright (B $< 13$) magnitudes.
\label{fig:optsrc}}

\figcaption[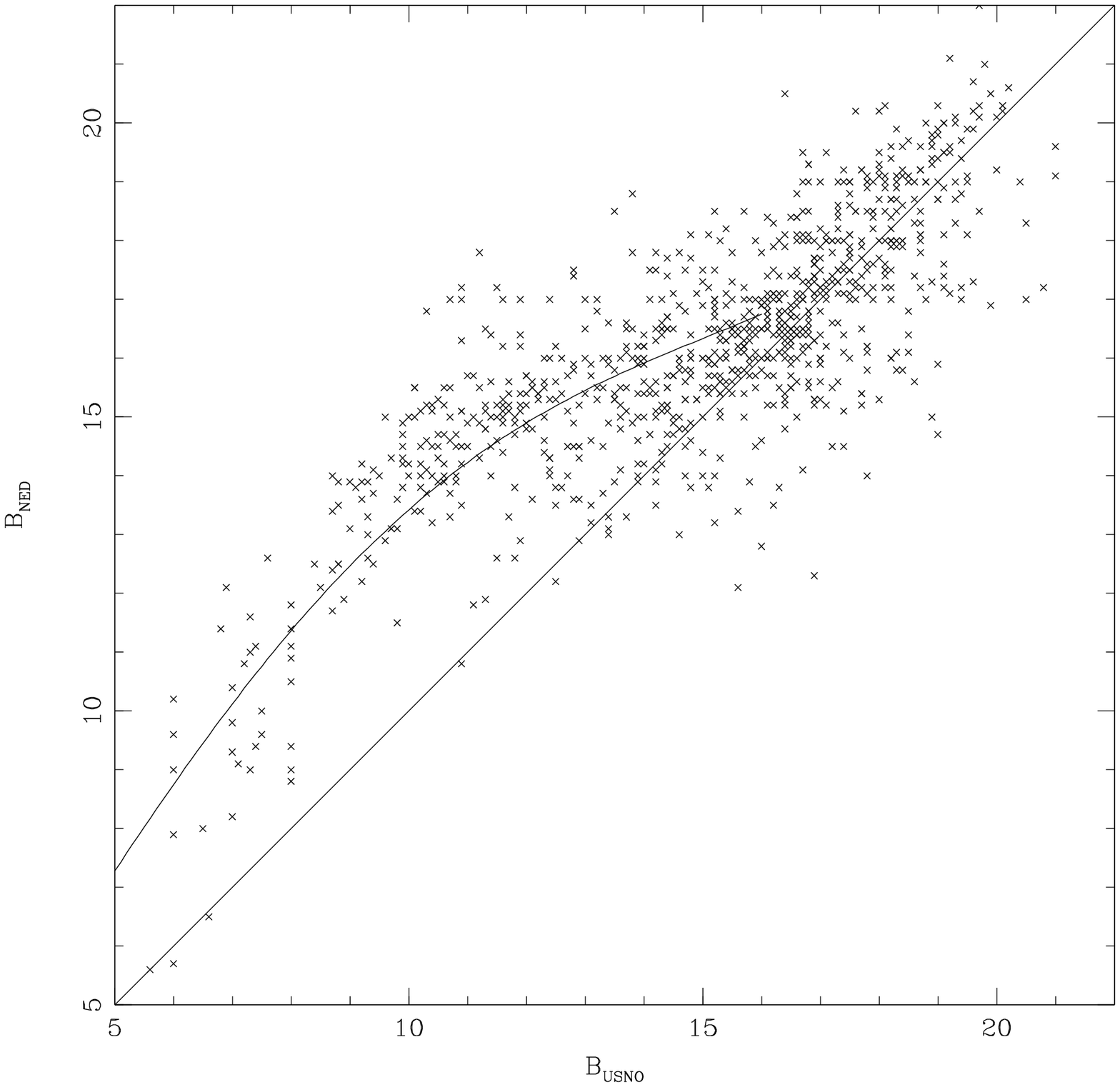]{
Comparison of B$_{\rm USNO}$ and B$_{\rm NED}$. The straight line
represents a one-to-one correspondence and the curve is a least-square 
fit to the data for B$_{\rm USNO}$ $<$ 16 mag.
\label{fig:mB_vs_mBned}}

\figcaption[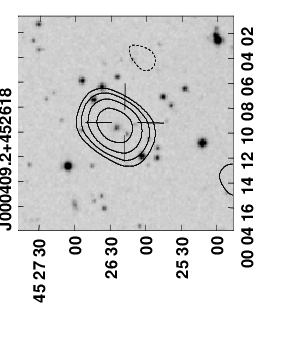]{
Sample finding chart of RBSC-NVSS sources. The NVSS radio contours
are overlaid on the DSS images, and the crosses denote the {\it ROSAT}
error box. In applicable fields, the 3 $\sigma$ error ellipses of IRAS
sources have been plotted.
\label{fig:charts}}

\figcaption[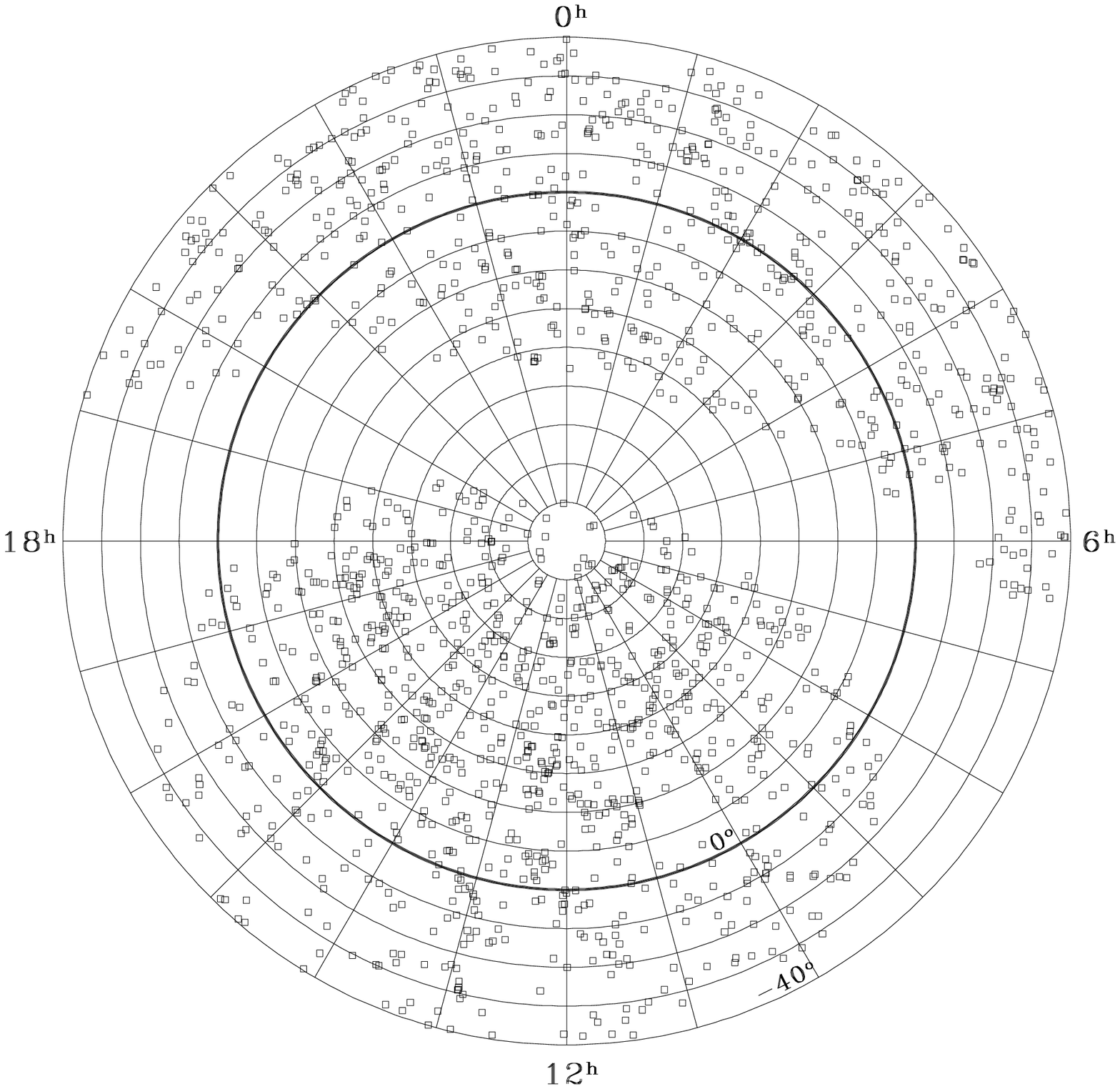]{
The sky distribution of the RBSC-NVSS sources. The overall
distribution is fairly uniform in the area $|b| > 15^\circ$, $\delta >
-40^\circ$.
\label{fig:skydist}}

\figcaption[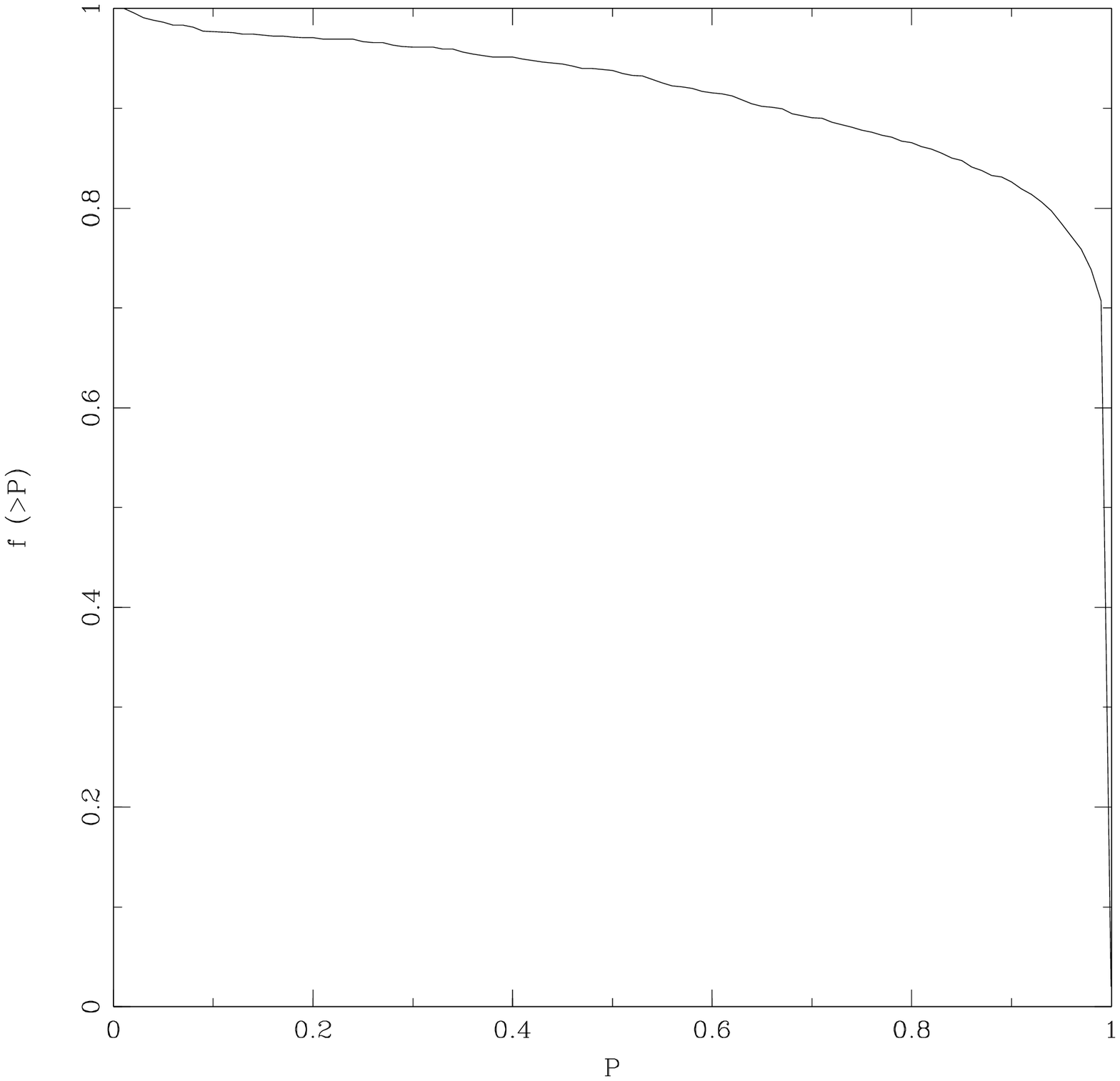]{
The cumulative fraction $f$ of RBSC-NVSS identifications as a function of
reliability $P$ (Equation~\ref{2dgenid}).
\label{fig:reliability}}

%
%	Figures
%

\begin{figure}
\plotone{f1.eps}
\end{figure}
\clearpage

\begin{figure}
\plotone{f2.eps}
\end{figure}
\clearpage

\begin{figure}
\plotone{f3.eps}
\end{figure}
\clearpage

\begin{figure}
\plotone{f4.eps}
\end{figure}
\clearpage

\begin{figure}
\epsscale{0.6}
\plotone{f5.ps}
\end{figure}
\clearpage

\begin{figure}
\plotone{f6.eps}
\end{figure}
\clearpage

\begin{figure}
\plotone{f7.eps}
\end{figure}

\clearpage

\end{document}